\documentclass[letterpaper,twocolumn,10pt]{article}
\usepackage{usenix,epsfig,endnotes}
\usepackage{color}
\usepackage[dvipsnames]{xcolor}
\usepackage{xurl}
\usepackage{xspace}
\usepackage{algorithm}
\usepackage{diagbox}
\usepackage[noend]{algpseudocode}
\usepackage{textcomp,upquote,listings, paralist}
\usepackage{enumitem}
\usepackage{amsmath}
\usepackage[font=small,labelfont=small,bf]{caption}
\usepackage[font=footnotesize,labelfont=footnotesize]{subcaption}
\usepackage{hyperref}
\usepackage{tabularx}
\usepackage{comment}
\usepackage{booktabs}
\usepackage{balance}
\usepackage{multirow}
\usepackage{makecell}
\usepackage{lastpage}
\usepackage{amssymb}
\usepackage{pifont}

\usepackage{tikz}
\usepackage{makecell}
\usepackage{multirow}
\usepackage{booktabs}
\newcommand*\circled[1]{\tikz[baseline=(char.base)]{
            \node[shape=circle,draw,inner sep=0.1pt] (char) {#1};}}

\newcommand{\ie}{i.e.\xspace}
\newcommand{\eg}{e.g.\xspace}

\newcommand{\para}{\noindent\textbf}
\newcommand{\paraue}[1]{\noindent\underline{\emph{#1}}}

\newcommand{\prepara}{\vspace{0.5em}}

\usepackage{titlesec}
\titlespacing*{\section}{0pt}{10pt plus 3pt minus 3pt}{5pt plus 3pt minus 2pt}
\titlespacing*{\subsection}{0pt}{5pt plus 3pt minus 2pt}{2pt plus 3pt minus 1pt}
\titlespacing*{\subsubsection}{0pt}{5pt plus 3pt minus 2pt}{2pt plus 3pt minus 1pt}
\titleformat{\section}{\large\bfseries}{\thesection}{1em}{}
\titleformat{\subsection}{\normalsize\bfseries}{\thesubsection}{1em}{}

\setlist[itemize]{noitemsep, topsep=0pt, leftmargin=*}
\setlist[enumerate]{noitemsep, topsep=0pt, leftmargin=*}

\newcommand{\sysname}{ShadowServe\xspace}
\newcommand{\shortsysname}{SS\xspace}

\begin{document}

\date{}

\title{\Large \bf \sysname: Interference-Free KV Cache Fetching for Distributed Prefix Caching}

{
\author{
\rm{Xingyu Xiang$^{\#}$ \enskip~~
    Raj Joshi$^{\#}$ \enskip~~
    Yuhan Liu$^{\dag}$ \enskip~~
    Jiayi Yao$^{\dag}$ \enskip~~
    Chenxingyu Zhao$^{\ddag}$
}
\vspace{1mm}
\\
\rm{Junchen Jiang$^{\dag}$ \enskip~~
    Yang Zhou$^{\P}$ \enskip~~
    Eddie Kohler$^{\#}$ \enskip~~
    Minlan Yu$^{\#}$}
\vspace{3mm}
\\
$^{\#}$Harvard University \enskip
$^{\dag}$University of Chicago \enskip
$^{\ddag}$University of Washington \enskip
$^{\P}$UC Davis
\vspace{5mm}
}

\maketitle}

\sloppy
%!TEX root = ./main.tex

\begin{abstract}

Distributed prefix caching accelerates long-context LLM serving by reusing KV cache entries for common context prefixes. However, KV cache fetches can become a bottleneck when network bandwidth is limited. Compression mitigates the bandwidth issue, but can degrade overall performance when decompression interferes with model computation.

We present \sysname, the first SmartNIC-accelerated, \emph{interference-free} prefix caching system for LLM serving. \sysname separates a control plane on the host and a data plane fully offloaded to the SmartNIC, which eliminates interference to both host GPU and CPU. To overcome the SmartNIC's limited compute and memory resources, we design a \emph{chunked pipeline} that parallelizes data plane operations across the SmartNIC's compute resources, and a \emph{minimal-copy memory management} scheme that reduces memory pressure on the SmartNIC. Compared to state-of-the-art solutions, \sysname achieves up to 2.2$\times$ lower loaded time-per-output-token (TPOT), and reduces time-to-first-token (TTFT) by up to 1.38$\times$ in low-bandwidth scenarios ($\le$\,20\,Gbps), translating to up to 1.35$\times$ higher throughput.

\end{abstract}

%!TEX root = main.tex
\section{Introduction}
\label{sec:intro}

Large Language Model (LLM) serving increasingly relies on longer contexts~\cite{longformer, unlimiformer, transformer-xl, routing-transformer, focused-transformer, memorizing-transformer} to improve generation quality and coherence, both in the form of longer input prompts~\cite{manus} and multi-turn interactions~\cite{beyond-single-turn}. These long contexts must be processed by the LLMs into intermediate states called the \emph{KV cache} before output generation begins. A key technique to reduce the high computational cost of this pre-processing step is \emph{prefix caching}, which can reuse precomputed KV cache data when different requests share prefixes in their contexts (\eg, system prompts~\cite{system-prompts}).

Prefix caches now commonly scale to distributed storage systems~\cite{cachegen, kdn, mooncake, deepseek-3fs, tokenlake}. There are three primary reasons for this shift. First, with production services generating petabytes of data daily, reusable KV cache entries far exceed the capacity of local GPU or CPU memory~\cite{mooncake}. Second, the lifespan of prefix cache entries can be long and unpredictable~\cite{kvcache-cache}, so high cache hit rates require large storage~\cite{cachegen}. Third, a distributed system allows KV cache entries to be shared across an entire cluster, allowing requests with identical prefixes to be routed to different nodes for load balancing~\cite{preble, mooncake, memserve}.

Distributed prefix caching is only beneficial when fetching the KV cache is faster than recomputing it. However, many LLM serving settings operate in relatively low-bandwidth environments, such as public cloud instances with capped network speeds~\cite{hexgen, spotserve, skyserve, heterogeneous-gpus, thunderserve, cachegen}, low-end GPUs that are typically paired with lower bandwidth~\cite{melange, helix, moe-lightening, heterogeneous-gpus, thunderserve}, and network-attached storage systems~\cite{nixl, dynamo-kv}.
In these environments, the network can become a bottleneck.

\begin{figure}[t]
    \centering
    \begin{subfigure}[b]{0.23\textwidth}
        \centering
        \includegraphics[width=\textwidth]{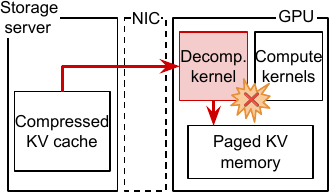}
        \caption{Prior work: KV cache fetching with GPU decompression.}
        \label{fig:gpu_dec}
    \end{subfigure}
    \hspace{0.2em}
    \begin{subfigure}[b]{0.22\textwidth}
        \centering
        \includegraphics[width=\textwidth]{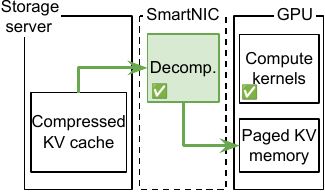}
        \caption{\sysname: KV cache fetching with SmartNIC decompression.}
        \label{fig:smartnic_dec}
    \end{subfigure}
    \caption{Solutions for distributed prefix caching.}
    \label{fig:motivation}
\end{figure}

Recent work~\cite{cachegen} observes that transferring \emph{compressed}, rather than raw, KV caches can mitigate this bottleneck and
improve the performance of prefix caching. KV caches are stored and transmitted in compressed format; serving nodes must
use their GPUs to decompress KV caches before using them (Figure~\ref{fig:gpu_dec}). Unfortunately, this introduces a new problem: \emph{interference with model computation}.
We find that concurrently running decompression and LLM serving on the same GPU causes both tasks to slow down, often by $\ge$\,30\% (\S\ref{sec:kv_cache_compression}). Offloading decompression to the host CPUs seems like a feasible workaround, but host CPUs in serving clusters are already burdened with other tasks like vector search and data preprocessing\cite{auncel, telerag, rago, pecan, tf-data, ray-data}, and they are inefficient at a variety of decompression algorithms (\S\ref{sec:smartnics}).

In this work, we present a new approach. \sysname (Figure~\ref{fig:smartnic_dec}) uses SmartNICs, rather than GPUs or CPUs, to fetch and decompress KV caches. SmartNICs are system-on-chip (SoC) devices with computation cores isolated from the host, integrating hardware accelerators for decompression. \sysname separates the prefix caching control plane from a SmartNIC-only data plane. The control plane, integrated into the LLM serving scheduler, initiates background fetches of KV caches as required. Meanwhile, the data plane handles all data-intensive operations, including decompression; since it runs on the SmartNIC, the host GPU is free to focus entirely on model computation.

However, using SmartNICs is not straightforward, as they are usually equipped with wimpy general-purpose on-chip cores and constrained memory subsystems. Naively offloading the entire data plane can create a new performance bottleneck.
We overcome these limitations with a \emph{chunked pipeline} for the SmartNIC data plane. Fetching is divided into four distinct stages: network fetching, lossless decompression, dequantization, and DMA to the GPU. Instead of processing an entire KV cache sequentially, we split it into fixed-size chunks that flow through these stages in a pipelined manner. This maximizes hardware utilization and improves operation throughput. To avoid resource contention, we use performance measurements to optimally partition the SmartNIC's compute resources, including the on-chip cores and dedicated accelerators, among these tasks.
In addition, we design a \emph{minimal-copy memory management} mechanism that eliminates redundant data copies and reduces memory access stalls; it works by pre-allocating and pinning all required memory buffers both on the SmartNIC and in GPU memory. Contention is avoided by giving each concurrent chunk in the pipeline its own dedicated memory region.

We implement a prototype of \sysname with the NVIDIA BlueField-3 DPU, a state-of-the-art SmartNIC, and evaluate it across a wide range of network bandwidth and output length settings. Compared to state-of-the-art GPU-decompression solutions, \sysname consistently achieves up to 2.2$\times$ lower loaded time-per-output-token (TPOT) across all settings, and reduces time-to-first-token (TTFT) by up to 1.38$\times$ in low-bandwidth scenarios ($\le$\,20\,Gbps). These advantages combined translate to up to 1.35$\times$ higher throughput. Our analysis also reveals that \sysname's performance is limited in high-bandwidth settings due to the SmartNIC's memory subsystem, identifying a key area for future hardware improvements.

We note that \sysname does not propose a new KV cache compression algorithm, but offloads existing ones from the host GPU/CPU to the SmartNIC. Therefore, we do not change any numerical property (\eg, accuracy or compression ratio) of the compression algorithms. \sysname's design is general to any compression algorithm that can be efficiently mapped to the hardware resources on the SmartNIC.

In summary, we make the following contributions.
\begin{itemize}
    \item We identify and characterize severe bidirectional interference between KV cache decompression and LLM model computation on the GPU.
    \item We present \sysname, the first prefix caching system to offload KV cache fetching and decompression to SmartNICs, with a clean control/data plane separation that eliminates host-side interference.
    \item We design a chunked pipeline and a minimal-copy memory management scheme to overcome the compute and memory limitations of current SmartNICs and maximize data plane throughput.
    \item We implement a prototype of \sysname, demonstrate that it outperforms state-of-the-art systems in terms of both latency and throughput in most settings, and analyze \sysname's bottleneck for the settings where it lags behind.
\end{itemize}

%!TEX root = main.tex
\section{Background and Motivation}
\label{sec:background}

\subsection{Prefix Caching}
\label{sec:prefix_caching}

LLM serving requests go through two phases, \emph{prefill} and \emph{decode}. In the prefill phase, the input prompt is processed to produce the intermediate states called the KV cache. Then, the decode phase utilizes the KV cache to autoregressively generate output tokens. Prefills are both resource- and time-consuming with complexity quadratic to the prompt length. Prefix caching~\cite{sglang, prompt-cache, claude-prompt-caching, openai-prompt-caching, deepseek-r1} helps requests skip the expensive prefill phase by reusing previously computed KV caches for repeated prompt prefixes.

Many existing systems \cite{vllm, sglang, pensieve, prompt-cache, chunkattention, infinigen, blendserve} cache KV data in local GPU/CPU memory, and some extend to local disk~\cite{cachedattention, ragcache, impress, cake, kvshare}, but on production-level services, prefix caching is expanding to distributed storage~\cite{cachegen, kdn, mooncake, deepseek-3fs, dynamo-kv}.
The expanded scale of remote storage supports the larger caches common on today's models---e.g., Llama-34B generates 19\,GB KV cache for a single 80K-token document~\cite{cachegen}, and production systems can generate PB-level KV caches~\cite{mooncake}---and allows KV caches to be held for longer intervals~\cite{cachegen}. Distributed storage also enables cross-node cache sharing, allowing requests with the same prefix to be routed to different machines for load balancing~\cite{preble, mooncake, memserve}.

\begin{figure}[t]
\centering

\begin{subfigure}[t]{0.45\textwidth}
    \includegraphics[width=\textwidth]{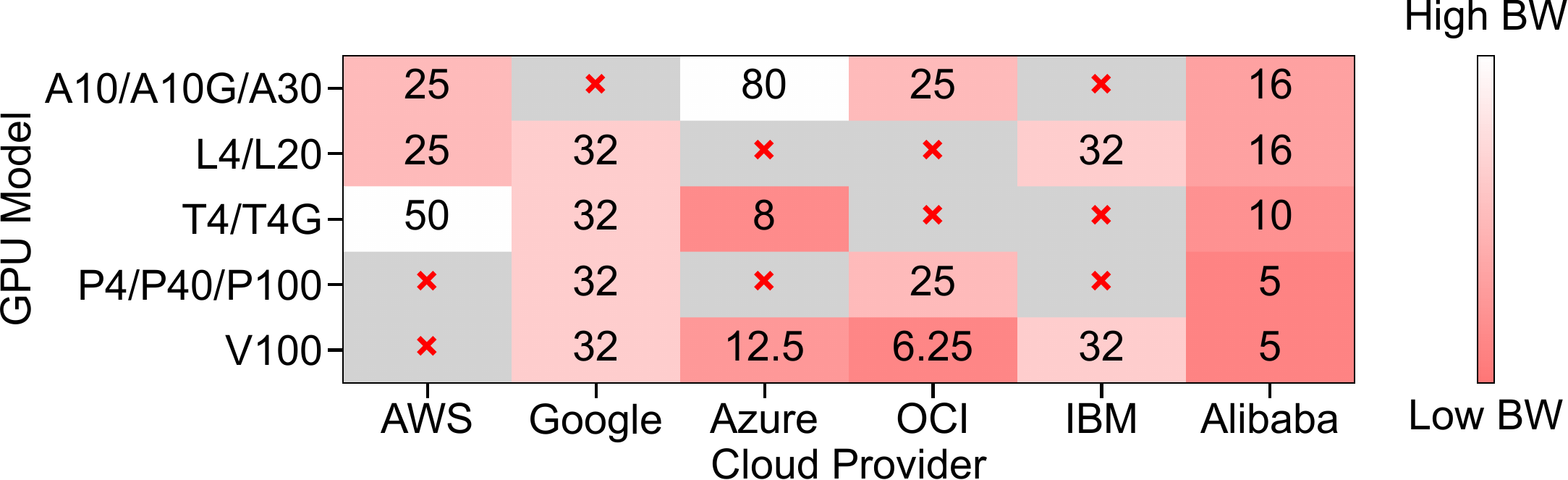}
\end{subfigure}

\caption{Maximum per-GPU network bandwidth (Gbps) for low-end GPU instances in major clouds. $\times$ = unsupported GPU model.}
\label{fig:gpu_bandwidth}
\end{figure}

Effective distributed prefix caching might seem to require high bandwidth. For example, Mooncake~\cite{mooncake} reports a sharp increase in TTFT and evident network congestion when inter-node bandwidth falls below 100\,Gbps; faster GPUs would push this breakeven bandwidth even higher.
Nevertheless, in some real-world scenarios, KV caches need to be fetched over limited bandwidth.
As serving extends to public cloud environments~\cite{hexgen, spotserve, skyserve, heterogeneous-gpus, thunderserve}, KV cache must be transmitted over cloud networks. Unlike specialized datacenter interconnects, cloud platforms typically cap inter-node bandwidth~\cite{skyplane, aws-instances, google-cloud-instances, oci-instances}. Low-end GPUs, nowadays a trendy choice for LLM serving\cite{melange, helix, moe-lightening, heterogeneous-gpus, thunderserve}, are provided with even lower bandwidth (Figure~\ref{fig:gpu_bandwidth}). Some works even transmit KV cache via the Internet over single-digit Gbps~\cite{kdn, cachegen}.
In addition, fetching KV cache from remote storage~\cite{nixl, dynamo-kv} further limits bandwidth due to disk access. In the AWS cloud, a GPU instance's access to network-attached storage like AWS Elastic Block Store can be limited to as low as 19\,Gbps~\cite{aws-instances}.

\subsection{KV Cache Compression}
\label{sec:kv_cache_compression}

CacheGen~\cite{cachegen} (now part of LMCache~\cite{lmcache}) shows that transmitting \emph{compressed} KV caches can significantly reduce bandwidth requirements, thus making distributed prefix caching valuable at lower network bandwidths. The KV cache is stored and fetched in a compressed form; it is only decompressed at the serving node where it is needed.

\begin{figure}[t]
    \centering
\hspace{3em}
    \begin{minipage}[t]{0.35\textwidth}{
        \begin{center}
        \includegraphics[width=\textwidth]{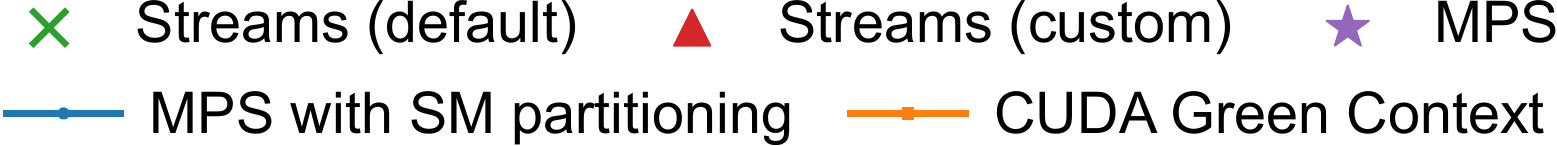}
        \end{center}}
    \end{minipage}
    \hfill
    \vspace{0.5em}
    
    \begin{subfigure}[b]{0.22\textwidth}
        \centering
        \includegraphics[width=\textwidth]{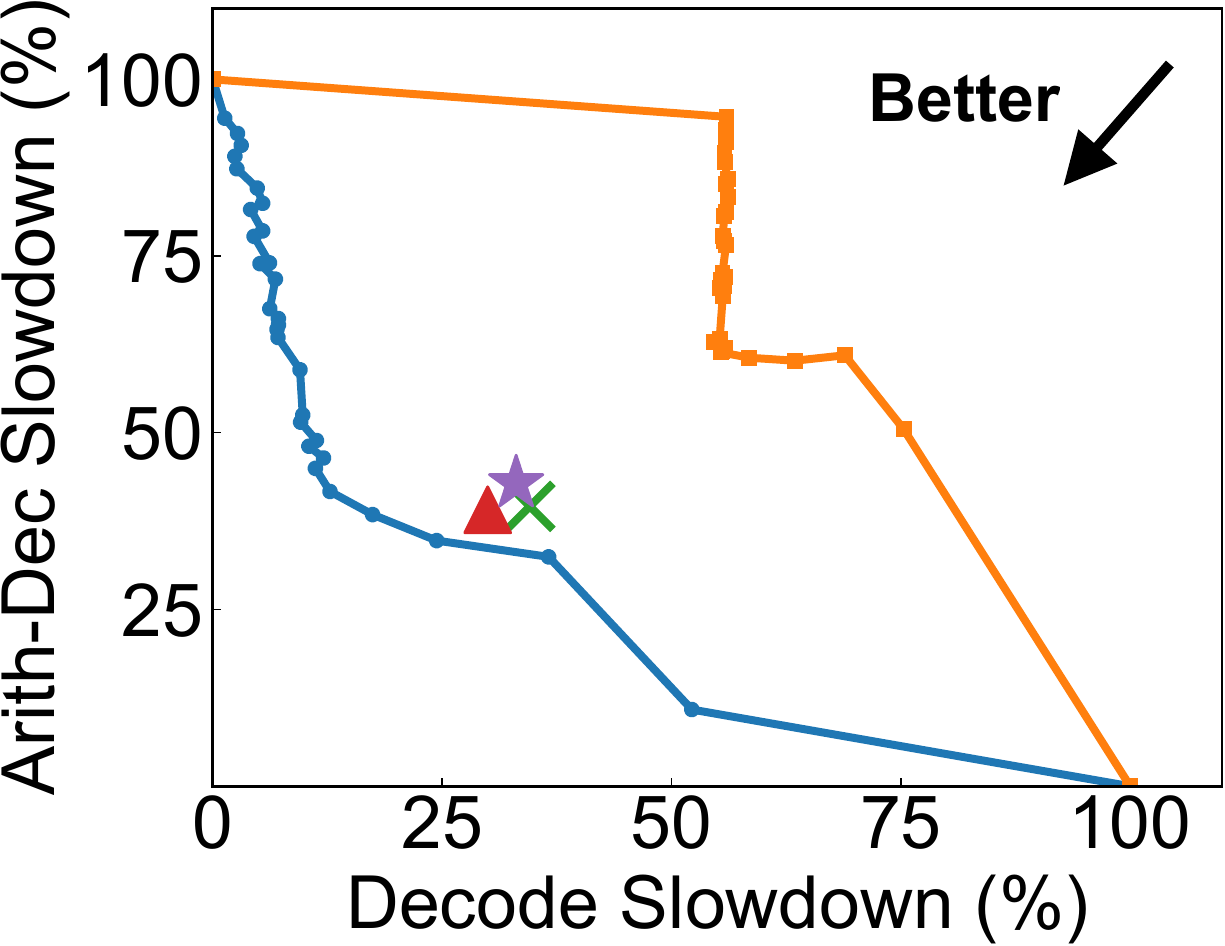}
        \caption{Interference between arithmetic decoding and LLM decode.}
        \label{fig:interference_cachegen}
    \end{subfigure}
    \hfill
    \begin{subfigure}[b]{0.22\textwidth}
        \centering
        \includegraphics[width=\textwidth]{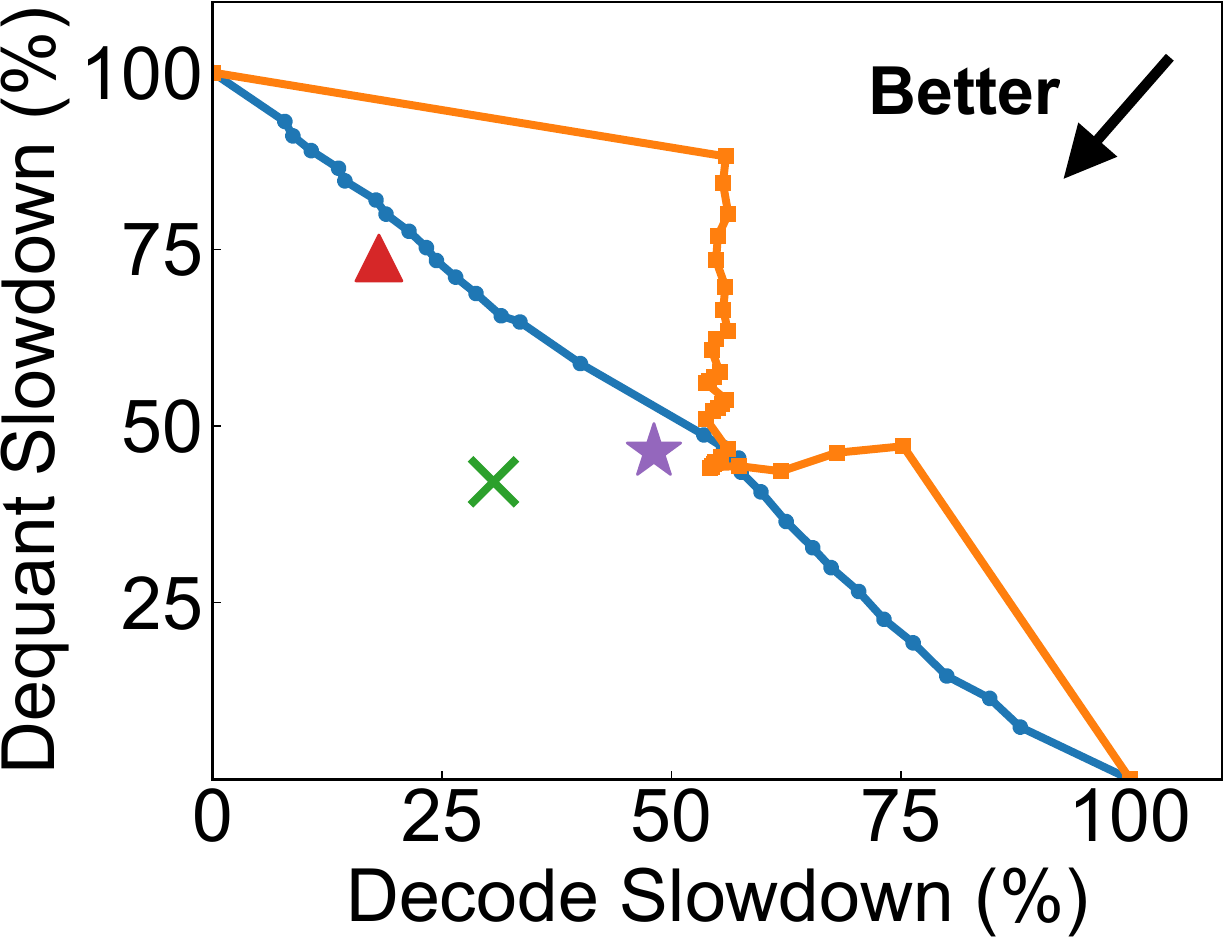}
        \caption{Interference between dequantization and LLM decode.}
        \label{fig:interference_dequant}
    \end{subfigure}
    \caption{Decompression and LLM decode interfere on GPU. No interference would correspond to 0\% slowdown for either task; we observe significant slowdowns across all GPU multitasking mechanisms. For CUDA streams, we launch LLM decode either in the default stream, or in a custom stream like decompression. MPS~\cite{mps} and Green Context~\cite{green-context} enable SM partitioning, and the two curves are plotted by assigning different numbers of SMs to the two tasks. Green Context performs poorly as it is an experimental feature and fails to partition memory bandwidth effectively~\cite{gpu-multitasking}.}
    \label{fig:interference}
\end{figure}

We note that this \emph{transmission-oriented} compression approach is distinct from \emph{runtime} KV cache compression, which aims to reduce the GPU memory footprint.
Transmission-oriented KV cache compression allows for more aggressive compression algorithms because it need not maintain a tensor format that the GPU can use directly. Therefore, powerful lossless compression like arithmetic coding~\cite{cachegen, arithmetic-coding} is used to further reduce the data transfer size beyond runtime methods like pruning~\cite{llmlingua, h2o, scissorhands} and quantization~\cite{kvquant, gear, kivi}.

\begin{table}[t!]
\begin{center}
{\small
\setlength{\tabcolsep}{8pt}
\begin{tabular}{ccc}
    \toprule
    \makecell{Algorithm}&\makecell{CPU (single core)}&\makecell{BF3 accelerator}\\\midrule
    \makecell{Deflate}
    &2.5&276.5\\
    \makecell{LZ4}
    &18.6&246.3\\
    \bottomrule
\end{tabular}}
\vspace{-1em}
\end{center}
\caption{Decompression output throughput (Gbps) for Deflate and LZ4, on 1 host CPU core and NVIDIA BlueField-3 DPU's decompression accelerator. We use the same KV cache data used in the final evaluation as input.}
\vspace{0.5em}
\label{tab:decompression_tput}
\end{table}

CacheGen delegates decompression to the GPU (Figure~\ref{fig:gpu_dec}). The compressed KV cache data is fetched to the GPU, decompressed using customized GPU kernels, and then stored into the GPU's KV cache memory.
Unfortunately, when decompression runs concurrently with model computation on the GPU, we observe severe bidirectional interference: both model serving performance and decompression throughput degrade substantially. 

To better characterize this interference, we run experiments that collocate decompression and LLM decode on an NVIDIA L40S GPU with different GPU multitasking mechanisms, and measure their respective slowdown (the throughput drop compared with running each operation on the entire GPU alone). We test both lossy and lossless compression mechanisms as used in CacheGen~\cite{cachegen}. The LLM decode has a context length of 30K; a minimal batch size (1 token) is used as it yields the least possible interference in long-context serving.

Figure~\ref{fig:interference} shows the results. Minimizing the slowdown of one task always leads to a drastic performance degradation of the other, regardless of the multitasking mechanism. For arithmetic decoding (Figure~\ref{fig:interference_cachegen}), it is not possible to limit both tasks' slowdown below 30\% at the same time. For dequantization (Figure~\ref{fig:interference_dequant}), the best mechanism still results in more than 25\% slowdown for both tasks. This fundamental bottleneck motivates moving decompression off the GPU.

\subsection{Opportunities with SmartNICs}
\label{sec:smartnics}

\begin{figure}[t]
\centering
\begin{minipage}[t!]{0.32\textwidth}
\includegraphics[width=\textwidth]{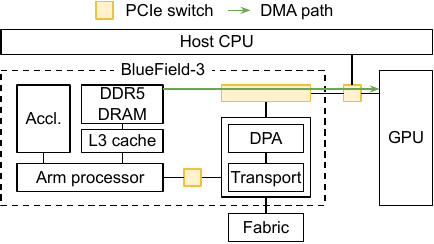}
\end{minipage}
\caption{Architecture of the NVIDIA BlueField-3 SmartNIC. It allows peer-to-peer DMA between the on-chip DRAM and GPU memory without host intervention.}
\label{fig:bf3}
\end{figure}

Our key idea is to offload KV cache fetching and decompression from GPU to a programmable I/O device, specifically a SmartNIC (Figure~\ref{fig:smartnic_dec}). This cleanly separates KV cache preparation and model computation by putting them on different hardware, thereby eliminating their interference.
For example, Figure~\ref{fig:bf3} shows the architecture of the NVIDIA BlueField-3 DPU~\cite{bf3} used in our system. It has a 16-core Arm Cortex-A78AE processor and 32GB DDR5 memory. It supports peer-to-peer DMA between the on-chip DRAM and the GPU memory, and it has a lossless decompression accelerator for both Deflate~\cite{deflate} and LZ4~\cite{lz4} algorithms directly accessible by the Arm processor.

The SmartNIC offers several advantages for fetching and decompression. It avoids using the GPU, which should spend as much time as possible on model serving. It avoids using the CPU, which is often burdened with auxiliary tasks (e.g., vector search in Retrieval-Augmented Generation~\cite{auncel, telerag, rago} or data preprocessing for multimodal models~\cite{pecan, tf-data, ray-data}), and has low decompression throughput per core (Table~\ref{tab:decompression_tput}). Data transfer overheads can be localized: on-chip DMA engines that facilitate peer-to-peer DMA between the SmartNIC and the GPU enable fetching directly into the GPU memory transparent to the host. Finally, SmartNICs yield superior KV cache fetching and decompression performance. SmartNICs are usually equipped with on-chip general-purpose CPU cores and dedicated lossless decompression accelerators that enable high-throughput processing on the KV cache data. 

\section{\sysname Overview}
\label{sec:overview}

\begin{figure}[t]
\centering
\begin{minipage}[t!]{0.4\textwidth}
\includegraphics[width=\textwidth]{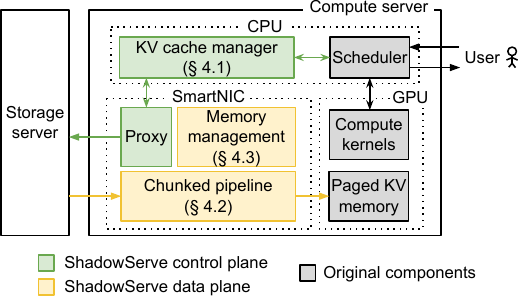}
\end{minipage}
\caption{\sysname overview.}
\label{fig:arch}
\end{figure}

We present \sysname, a SmartNIC-accelerated prefix caching system that supports \emph{interference-free} KV cache fetching. At the core of \sysname is a clean separation between the \emph{control plane} and the \emph{data plane} of prefix caching. The control plane runs on the host CPU and schedules KV cache fetches, and the data plane, fully offloaded to the SmartNIC, handles individual network I/O and decompression operations. By offloading only the data operations to the SmartNIC, \sysname eliminates host-side interference while keeping the SmartNIC stack simple to implement.

Fully unleashing the benefits of SmartNIC offloading requires addressing two challenges on the SmartNIC: limited compute and memory resources. On the compute side, SmartNICs integrate general-purpose on-chip cores that are significantly less powerful than standard server CPUs, and performing data-intensive operations naively on these cores can create a new performance bottleneck. On the memory side, the SmartNIC's memory subsystem is also restricted. For example, the BlueField-3 has a small cache hierarchy of 1\,MB, 8\,MB, and 16\,MB L1, L2, and L3 caches~\cite{bf3}, which must serve all pipelined operations. Although the on-chip DDR5 DRAM provides a high bandwidth of >\,80\,GB/s, there are only two memory channels. For comparison, a typical host memory system provides eight memory channels per CPU socket. Meanwhile, most of the offloaded data operations are memory intensive.

To address the compute challenge and maximize throughput on the SmartNIC, we design the data plane as a \emph{chunked pipeline} (\S\ref{sec:chunked_pipeline}). This design breaks the end-to-end fetching process into distinct stages, and assigns them to different SmartNIC resources (\eg, CPU cores and hardware accelerators), to enable maximum parallelism and resource utilization. To overcome the memory challenge, we design a \emph{minimal-copy memory management} scheme (\S\ref{sec:memory_management}) for the data plane. This mechanism carefully manages on-chip memory buffers to ensure a smooth flow of data through the pipeline, avoiding expensive memory registration and data copies between stages.

Figure \ref{fig:arch} provides an overview of the \sysname architecture, built upon the separation of control and data planes. \sysname integrates well with existing serving framework components, including a scheduler that accepts user requests arriving at the system and dispatches compute kernels to the GPU, and a paged KV memory region in the GPU memory. The core prefix caching functionality is implemented in our two planes.

\prepara
\para{Control plane.}
The control plane manages the orchestration of KV cache fetches. It enables fully \emph{asynchronous} fetching that moves KV cache fetching off the critical path (\S\ref{sec:asynchronous_fetching}) and is composed of two major components.

\paraue{KV cache manager.}
The KV cache manager runs on the host CPU. It communicates with the scheduler each iteration to identify requests eligible for KV cache fetching. It temporarily takes ownership of these requests and sends their metadata to the SmartNIC proxy for fetching. Once the data plane has fetched the KV cache, the manager is notified, and it seamlessly submits the ready requests back to the scheduler to begin token generation. This coordination occurs in the background, transparent to the serving scheduler (\S\ref{sec:asynchronous_fetching}).

\paraue{SmartNIC proxy.}
The proxy runs on the SmartNIC on-chip cores and communicates with the KV cache manager via a communication channel. Upon receiving fetch requests, it communicates request metadata with the remote storage server and instructs the data plane to begin the fetching process. It oversees the entire operation on the SmartNIC and notifies the KV cache manager upon successful completion, \ie, when the KV cache is ready in the GPU memory.

\prepara
\para{Data plane.}
The data plane resides entirely on the SmartNIC and is responsible for efficiently retrieving, decompressing, and transferring KV cache data from the remote storage server to GPU memory. The data plane's key components are the \emph{chunked pipeline} (\S\ref{sec:chunked_pipeline}), which maximizes resource utilization, and our \emph{minimal-copy memory management} mechanism (\S\ref{sec:memory_management}), which alleviates the on-chip memory bottleneck. By offloading the entire data path from the host, \sysname eliminates GPU interference while achieving high throughput.

\prepara
While our discussion focuses on a single GPU-SmartNIC pair, \sysname's design applies to multi-GPU serving environments that leverage tensor and pipeline parallelism~\cite{megatron-lm, alpaserve}. This extension requires pairing each GPU with a dedicated SmartNIC, where each SmartNIC's data plane independently fetches the KV cache portion for its associated GPU. We believe this one-to-one architecture is practical, as datacenter machines often equip each GPU with a dedicated high-performance NIC already~\cite{dgx, grand-teton, dell, google-arch}.
%!TEX root = main.tex
\section{\sysname Design}
\label{sec:design}

\subsection{Asynchronous Fetching}
\label{sec:asynchronous_fetching}

Asynchronous fetching is essential for realizing the full benefits of prefix caching in a system with remote storage. Fetching a request's KV cache from a storage server introduces I/O latency that, if handled synchronously, would stall the GPU and severely limit system throughput. The primary goal of our asynchronous design is to decouple this I/O-bound fetching process from the main execution flow. By fetching KV caches for some requests in the background while the GPU processes others, we can hide the I/O latency, maximize hardware utilization, and maintain high serving throughput.

To achieve asynchronous fetching, we design a \emph{KV cache manager} running in parallel with the serving scheduler in the host CPU. The manager interacts with the scheduler, transparently moving eligible requests out of the execution flow and fetching their KV cache in the background.

\begin{figure}[t]
\centering
\begin{minipage}[t!]{0.4\textwidth}
\includegraphics[width=\textwidth]{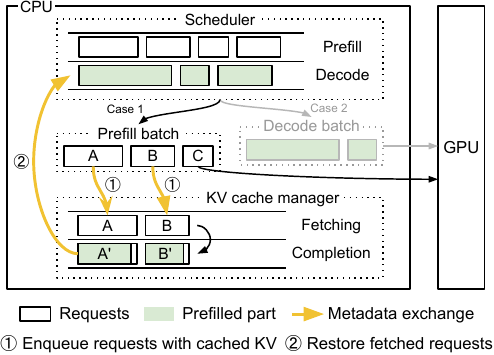}
\end{minipage}
\caption{\sysname's asynchronous fetching.}
\label{fig:async_fetching}
\end{figure}

The KV cache manager keeps track of requests undergoing KV cache fetching by maintaining two internal queues, \emph{fetching} and \emph{completion}, shown in Figure~\ref{fig:async_fetching}. The \emph{fetching} queue contains requests eligible for fetching, and the \emph{completion} queue holds requests that have completed the fetching process and are ready to resume execution. In our current design, we use simple FIFO queues; the manager processes fetch jobs serially. However, because the number of tokens in each fetch job is known, the fetching times can be estimated, opening the door to more advanced scheduling (\eg, Shortest Job First), which we leave as future work.

\prepara
\para{Batch interception.}
The KV cache manager identifies and moves eligible requests to the fetching queue without interrupting the scheduler. It achieves this with a non-intrusive \emph{batch interception} mechanism that operates in lockstep with the scheduler. In each iteration that the scheduler produces a prefill batch, the manager performs a two-way exchange. First, it intercepts the batch and strips out any requests eligible for remote KV cache fetching, moving them to its fetching queue to be processed in the background (\circled{1} in Figure~\ref{fig:async_fetching}). Note that the KV cache manager only intercepts prefill batches, as decode requests already have their KV cache computed. Simultaneously, it checks its completion queue for any requests that have finished fetching from a previous iteration and restores them to the scheduler for execution (\circled{2} in Figure~\ref{fig:async_fetching}). These two operations are atomic from the scheduler's perspective. The scheduler then proceeds with its (now modified) batch, unblocked by the KV cache fetches happening asynchronously.

For example, in Figure~\ref{fig:async_fetching}, the scheduler either produces a prefill batch (case 1) or a decode batch (case 2) for the current iteration. For case 1, the KV cache manager examines the prefill batch, consisting of requests A, B, and C, to see if any of them are eligible for KV cache reuse. Here, request C's KV cache is not stored in the storage server, so its metadata is parsed and sent to GPU as usual. For requests A and B, their KV cache is stored, so they are temporarily removed from the current batch and the scheduler, and added to the manager's fetching queue. Meanwhile, any requests in the completion queue are restored to the scheduler.

A natural question is why not begin fetching KV cache earlier, \eg, when the requests \emph{arrive}, instead of when they \emph{are scheduled}. The reason is that for every new request, its entire prompt's KV cache space in the GPU is allocated on demand when it is scheduled for the first time as a prefill request, and we can only begin fetching after this GPU memory allocation. This lazy allocation strategy aims to lower GPU memory consumption. Without it, a burst of requests could easily exhaust GPU memory.

\prepara
\para{Background fetching.}
After identifying eligible requests and moving them to the fetching queue, the KV cache manager needs to tell the data plane to fetch the KV cache for these requests in the background. It does so by running a \emph{background fetching loop}. Each time, it retrieves a request from the fetching queue, fetches the request's KV cache into the paged memory via the data plane, and puts the request into the completion queue. In Figure~\ref{fig:async_fetching}, requests A and B are submitted to the data plane, and put to the completion queue after fetching. Again, this process is transparent to the scheduler, which continues forming and dispatching batches without interruption.

When a request is put into the completion queue, a straw-man solution would be to update it as fully prefilled. However, this will cause a bug because only populating the KV cache is not equivalent to a full prefill. The reason is that, a full prefill additionally produces the \emph{first output token} by sampling from the \emph{hidden states}, \ie the computation result of the model, while the KV cache is just the \emph{intermediate state} of this computation. This problem is also present in previous prefix caching systems~\cite{cachegen}, and some propose storing both the KV cache and the hidden states in the storage server. In our system, we adopt a similar workaround to CacheGen~\cite{cachegen} by marking the last token of the request as not yet prefilled, illustrated as A' and B' in Figure~\ref{fig:async_fetching}. When these last-token-prefill jobs are restored to the scheduler, they are piggybacked immediately in the next batch.

\prepara
\para{Limitations.}
Currently, our design does not support chunked prefill. We also do not support partial hits, so a user request will either have its entire KV cache stored in the storage server, or have no KV cache stored at all. Nevertheless, we believe these features to be orthogonal to our contribution, and easy to integrate with our design. We discuss how to support them in \S\ref{sec:discussion}.

\subsection{Chunked Pipeline}
\label{sec:chunked_pipeline}

\begin{figure}[t]
\centering
\begin{minipage}[t!]{0.35\textwidth}
\includegraphics[width=\textwidth]{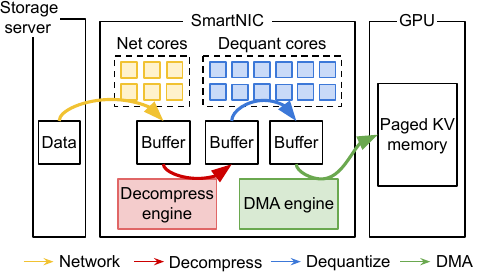}
\end{minipage}
\caption{\sysname's chunked pipeline.}
\label{fig:chunked_pipeline}
\end{figure}

To address the SmartNIC compute bottleneck, we divide the data operations into four stages and explicitly allocate SmartNIC compute resources across them, including CPU cores and accelerators. To maximize SmartNIC resource utilization, we further pipeline these stages to make them run in parallel.

We perform four data operations in total on the SmartNIC, shown in Figure~\ref{fig:chunked_pipeline}. First, we need to fetch the compressed data from the remote storage server via the network. Then, we need to perform lossless decompression on the data, followed by dequantization, to get the original KV cache tensor. Finally, we need to copy the original KV data from the on-chip memory to the GPU via DMA. The reason we need both decompression and dequantization is that the two steps combined ensure maximum compression ratio over the high-entropy KV cache data, following prior work~\cite{cachegen}.

\prepara
\para{Resource partitioning.}
To ease the load on the wimpy Arm cores, we first offload the lossless decompression and DMA operations to the hardware accelerators on the SmartNIC (see Figure~\ref{fig:chunked_pipeline}). Decompression accelerators and DMA engines are present in many SmartNIC SoCs~\cite{iotcp, bf3, intel-ipu, amd-pensando}. For example, BlueField-3 supports hardware acceleration for lossless decompression with high throughput~\cite{doca-compress}, and it has an on-chip DMA engine capable of copying data from the on-chip DRAM directly into host GPU via peer-to-peer DMA over PCIe (\S\ref{sec:kv_cache_compression}). Both engines work fully asynchronously to the Arm cores and incur negligible CPU load.

For the network and dequantization operations, we partition the on-chip Arm cores into two parts, assigning one part to each operation (see Figure~\ref{fig:chunked_pipeline}). We further parallelize network transmission and dequantization to their assigned cores. For network, we split the transmitted data into equal-sized slices and receive each slice with one core. For dequantization, we use vector-wise data binning, and we split the KV cache tensor into 1D vectors, distributing them equally among the assigned cores. The number of cores assigned to each part is determined by the network and dequantization throughput per core to ensure balanced throughput.

To save context switching overhead, we start and pin all threads to the assigned core during SmartNIC program initialization, and use a lightweight thread-safe FIFO queue to pass tasks from the main program to these worker threads.

\prepara
\para{Four-stage chunked pipeline.}
To minimize fetch latency and maximize resource utilization, we pipeline all the four operations on the SmartNIC. Instead of processing each request's entire KV cache sequentially through the four stages, we split each KV cache tensor into fixed-size chunks, allowing each chunk to flow independently through the pipeline. This overlaps network transmission, lossless decompression, dequantization, and DMA transfer on the SmartNIC. Since we partitioned the compute resources, the operations have minimal compute interference with each other. This chunked pipeline enables continuous full utilization of all resources on the SmartNIC, and reduces end-to-end latency to the slowest of the four stages.

Prior systems also have similar pipelining techniques. For example, CacheGen~\cite{cachegen} pipelines network transmission and GPU decompression. However, both CacheGen's lossless decompression and dequantization utilize the entire GPU, so they cannot be pipelined. Meanwhile, \sysname assigns different resources to all operations, enabling more fine-grained pipelining and parallelism.

\subsection{Memory Management}
\label{sec:memory_management}

\begin{figure}[t]
\centering
\begin{minipage}[t!]{0.47\textwidth}
\includegraphics[width=\textwidth]{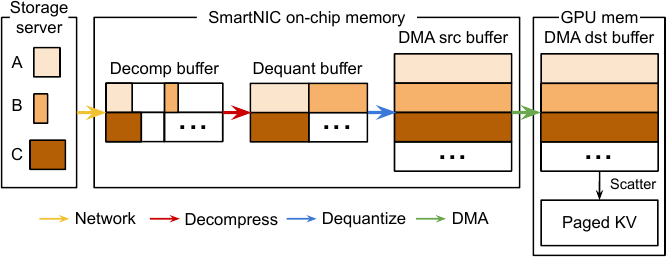}
\end{minipage}
\caption{\sysname's memory management. It is also possible to perform DMA directly into paged KV memory, but this results in scattered data copies which significantly degrade DMA throughput.}
\label{fig:mem_management}
\end{figure}

To address the SmartNIC memory bottleneck, we design a \emph{minimal-copy} memory management mechanism that eliminates redundant copies (Figure~\ref{fig:mem_management}). It integrates seamlessly with the SmartNIC programming model and the chunked pipeline.

\prepara
\para{Memory pinning.}
Instead of creating memory buffers during runtime, we pre-allocate and pin all buffers during program initialization. They include (1) the decompression buffer, the dequantization buffer, and the DMA source buffer in the SmartNIC on-chip memory; (2) the DMA destination buffer in the GPU memory. Only these memory regions are accessed by the operations. For example, the DMA destination buffer serves as the sole endpoint for DMA transfers into the GPU.

This technique has two benefits. First, it minimizes memory copies on the SmartNIC, as a subsequent operation directly reads from the previous one's output buffer. For example, in Figure~\ref{fig:mem_management}, dequantization can be directly carried out on the output of decompression, \ie, data in the dequantization buffer, without any data copy. Second, it reduces memory registration overhead. All memory regions accessed by hardware accelerators should be registered with the SmartNIC, and memory registration during runtime usually causes significant delay (\eg, up to 3$\times$ fetching latency on BlueField-3). Similar memory pinning strategies are present in other systems like NCCL~\cite{nccl}.

A straw-man solution here is to reserve larger buffers so that each request's entire KV cache can fit in the buffers. However, this solution results in high GPU memory consumption caused by the DMA destination buffer, which leads to memory contention with model serving and even out-of-memory errors on the GPU. Therefore, we only allocate a medium-sized buffer to ensure a bounded GPU memory footprint. If one request's KV cache data is larger than any of the buffers, we perform the KV cache transfer for such a request in multiple \emph{rounds}, each time fetching a subset of chunks that fit within the available buffer. 

\prepara
\para{Buffer partitioning.}
With the chunked pipeline design (\S\ref{sec:chunked_pipeline}), each chunk goes through the pipeline independently, and operations on different chunks need to access the same memory buffer at the same time. For example, in Figure~\ref{fig:mem_management}, we fetch 3 compressed KV cache chunks---A, B, and C---in the current round. When A is being dequantized, B is being decompressed in parallel, and both operations need to access the dequantization buffer. To avoid memory contention, we need to partition the buffers across chunks.

The challenge is to decide the memory size allocated to each chunk in each buffer, which we call \emph{occupancy}. Luckily, for a given chunk, its occupancy in most of the buffers can be neatly determined. First, both in the DMA source buffer and the DMA destination buffer, a chunk's occupancy is its original KV cache tensor size, which is determined by the number of tokens in the chunk and the model dimensions (\eg, number of layers and channels). This size can be calculated locally during runtime. Since our quantization algorithm exactly halves the original data size, the occupancy in the dequantization buffer is half of the occupancy in the DMA buffers.

The last difficulty is to determine the occupancy in the decompression buffer, as we do not know the compressed data size for each chunk. A straw-man solution is to query the storage server, but it adds querying overhead for each round. Our solution leverages the fact that compressed size is always smaller than the original, and directly uses the same occupancy in the dequantization buffer as the decompression buffer. Since compressed size is smaller, we create some unused fragments in the decompression buffer, illustrated in Figure~\ref{fig:mem_management}. With this strategy, each chunk's occupancy in both the decompression and dequantization buffers will always be half of that in the DMA buffers. Based on this observation, we make the decompression and dequantization buffers also half the size of the DMA buffers, so that for each round, the buffers are able to fit the same set of chunks.

Occupancies are computed with negligible overhead for every chunk in a round before the actual fetching begins. After that, the buffers are logically partitioned according to the computed occupancies, so that each chunk only utilizes its assigned part in all buffers. For example, in Figure~\ref{fig:mem_management}, assuming one row in each buffer corresponds to 4\,MB of data, each chunk's occupancy is 2\,MB in the decompression and dequantization buffers, and 4\,MB in the DMA buffers. When, for instance, chunk B is doing dequantization, it reads from its own allocated region in the dequantization buffer (2\,MB), and writes to its allocated region in the DMA source buffer (4\,MB).

\prepara
\para{Scattering.}
Every round after fetching a group of chunks, we launch a lightweight kernel to \emph{scatter} the contiguous KV cache tensor in the DMA destination buffer into the paged KV memory. As both regions are in the GPU, this memory copy kernel causes negligible overhead, which is further mitigated by the fact that we launch this kernel every \emph{round} after fetching multiple chunks, while previous works launch a decompression kernel for every \emph{chunk}.

%!TEX root = main.tex
\section{Implementation}
\label{sec:implementation}

We implement a system prototype of \sysname with 7K lines of code in Python and C++, and integrate it with vLLM, a state-of-the-art LLM serving system. \sysname can be integrated with any serving framework, and we choose vLLM because it is the most widely used. Our implementation leverages the NVIDIA BlueField-3 DPU, but \sysname can be implemented on any SmartNIC with decompression accelerators and P2P DMA capability, present on most current SmartNICs like Intel IPU~\cite{intel-ipu} and AMD Pensando~\cite{amd-pensando}.

\prepara
\para{Host side.}
The KV cache manager runs as a thread in the vLLM process, communicating with the SmartNIC proxy via DOCA Comch~\cite{doca-comch}, an event-driven message-based channel over PCIe. All SmartNIC-related functionalities are exposed to the manager as a C++ pybind function. This function communicates with the SmartNIC proxy, and blocks until it receives completion notification from the proxy. We release the Python Global Interpreter Lock (GIL) inside the pybind function, so that the vLLM scheduler thread is not blocked.

\prepara
\para{SmartNIC side.}
We use TCP as the transport between the SmartNIC and the storage server, and utilize NVIDIA Accelerated IO (XLIO) library, a user-space TCP/IP stack compatible with POSIX socket APIs, to improve network performance. We disable Nagle's algorithm and TCP delayed acknowledgment to avoid stalls caused by small metadata exchanges between the SmartNIC proxy and the storage server.

For the chunked pipeline, we set the chunk size to 256 tokens, following prior work~\cite{cachegen}. We allocate 2 of the 16 Arm cores on BlueField-3 to network (XLIO TCP), and the remaining 14 cores to dequantization. We further use SIMD (Arm Neon~\cite{arm-neon}) to boost the dequantization performance on the Arm cores. Both the decompression and DMA accelerators on BlueField-3 require polling from the Arm cores, and we set the polling interval as 10$\mu$s, which ensures responsiveness while incurring negligible load on the Arm cores.

For memory management, we allocate 0.5\,GiB for both the DMA source and destination buffers, and 0.25\,GiB for the decompression and dequantization buffers. In addition to the DMA destination buffer, we reserve an additional 0.5\,GiB in the GPU for tensor reshaping and scattering, capping our GPU memory footprint at 1\,GiB. This size can be further reduced at the cost of some additional overhead, caused by launching the scattering kernel for more fetching rounds (\S\ref{sec:memory_management}). For scattering, we use the \texttt{reshape\_and\_cache} CUDA kernel implemented by vLLM. To handle BlueField-3's 2\,MiB limit for accelerator operations (e.g., decompression and DMA), we pre-slice data into compatible blocks during the initial compression stage to avoid splitting already-compressed data.

\prepara
\para{Compression algorithm.}
We use vector-wise data binning for quantization, following previous work~\cite{cachegen}. For each vector in the KV cache tensor, this method finds the maximum absolute value and uses it to scale all elements into a predefined set of bins. The resulting quantized data is losslessly compressed using the Deflate algorithm. We chose Deflate because it is hardware-accelerated by BlueField-3 and achieves a superior compression ratio on the quantized KV cache data compared to LZ4.

\prepara
\para{Storage server.} The remote storage server is organized as a key-value store, where each entry stores the compressed KV cache for a given chunk. The key is the prefix hash of the request's prompt up to the chunk in question. The KV cache manager checks whether a request's KV cache is stored in the storage server by querying whether the last chunk's prefix hash exists as a key in the server.

%!TEX root = main.tex
\section{Evaluation}
\label{sec:evaluation}

This section aims to answer the following questions: 
\begin{enumerate}
    \item What is the throughput and latency of \sysname under different settings compared to existing approaches, and what is the trade-off space (\S\ref{sec:end_to_end})?
    \item How is the SmartNIC data plane performing, and what is the bottleneck on the SmartNIC (\S\ref{sec:smartnic_perf})?
    \item What are the effects of asynchronous fetching, chunked pipeline, and memory management on \sysname's performance (\S\ref{sec:ablation_studies})?
\end{enumerate}

\subsection{Experiment Setup}
\label{sec:experiment_setup}

\para{Testbed.}
Our testbed consists of a single machine with two 16-core Intel Xeon Gold 6526Y CPUs at 3.50\,GHz and 504\,GB memory. It has an NVIDIA L40S GPU and an NVIDIA BlueField-3 DPU under NUMA node 0, and a dual-port Mellanox ConnectX-7 400\,Gbps NIC under NUMA node 1, all connected via PCIe 4.0 $\times$16. The host machine runs Ubuntu 24.04 (kernel v6.8.0), and the BlueField-3 runs Ubuntu 22.04 (kernel v5.15.0) and DOCA v2.9.0. The BlueField-3 features a 16-core Cortex-A78AE Arm CPU with 32\,GB of on-chip memory, and integrates a dual-port Mellanox ConnectX-7 400\,Gbps NIC. One port of the BlueField-3 and the CX7 NIC under NUMA node 1 are connected via a loopback link. The CX7 NIC is rate limited with Mellanox OFED QoS~\cite{mlnx-qos}. In our experiments, the LLM serving engine runs on NUMA node 0, and the KV cache storage server runs in a separate network namespace on NUMA node 1, accessible through the physical loopback link between the two NICs.

\prepara
\para{Baselines.}
We compare \sysname to two baseline LLM serving systems.
\begin{itemize}
    \item \textbf{vLLM~\cite{vllm}.} It is a popular LLM inference system. We use vLLM v0.8.1, the latest version when starting the project.
    \item \textbf{CacheGen-Async}. It is a state-of-the-art prefix caching system with asynchronous KV cache fetching. Because the original implementation of CacheGen~\cite{cachegen} does not support asynchronous fetching, we enable it with our design (\S\ref{sec:asynchronous_fetching}). It bypasses the BlueField-3, using it as a normal NIC, and utilizes GPU for KV cache decompression. Like \sysname, it uses TCP as transport with the same optimizations discussed in \S\ref{sec:implementation}, and implements a chunked pipeline of network transmission and decompression.
\end{itemize}

For CacheGen-Async, we use one custom CUDA stream for model computation, and another for KV cache decompression. \sysname also needs to run a lightweight scattering kernel in the GPU (\S\ref{sec:memory_management}), and we put it in a custom stream as well. We also evaluate the effect of using the default stream for model computation in \S\ref{sec:more_baselines}.

\prepara
\para{Methodology.}
To measure the pure remote fetching performance of \sysname and CacheGen-Async, we configure the experiment to ensure every request triggers a fetch from the remote storage server, similar to CacheGen~\cite{cachegen}. This is achieved by two means: First, we disable their local prefix caching (in GPU/CPU memory). Second, we ensure a 100\% cache hit rate on the remote server, so that requests never need to be recomputed. To facilitate this, all necessary KV cache data is pre-loaded into the storage server's memory before each experiment.
Conversely, vLLM serves as our recomputation baseline. We force it to recompute the KV cache for every request by also disabling its local prefix caching.

\prepara
\para{Workloads.}
We evaluate \sysname on two models of different sizes, Llama-8B and Mistral-7B. Both models are fine-tuned to take long contexts (32K for Mistral-7B and 128K for Llama-8B). We use two different datasets, TriviaQA and NarrativeQA. Both datasets are part of the LongBench~\cite{longbench} benchmark suite, designed to test the reading comprehension ability of LLMs by giving the LLMs a story or script, provided as a long document, and asking questions on it. TriviaQA has a medium length of 9.3K tokens, and NarrativeQA has a medium length of 14K. Both datasets have P95 lengths of 15K.
For each experiment, we randomly sample 200 requests from one dataset, and the request arrival timestamps follow the Poisson process with different average rates.

\prepara
\para{Metrics.}
We measure two primary latency metrics for LLM serving: TTFT (time-to-first-token) and TPOT (time-per-output-token). For a given request, TTFT measures the latency of generating the first output token from the moment a request arrives in the system. For systems without prefix caching (vLLM), unloaded TTFT indicates the latency of prefill computation, while for prefix caching systems (CacheGen-Async and \sysname), it indicates the latency of fetching and decompressing the KV cache. TPOT measures the interval between the generation of consecutive output tokens of a request. For asynchronous fetching systems, a higher TPOT indicates stronger interference between KV cache decompression and model computation.

\subsection{End-to-End Performance}
\label{sec:end_to_end}

\subsubsection{Overall Performance}

\begin{figure}[t]
\centering
\includegraphics[width=0.32\textwidth]{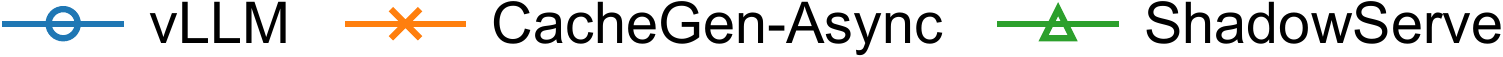}
\vspace{0.2em}

\begin{subfigure}[t]{0.237\textwidth}
    \includegraphics[width=\textwidth]{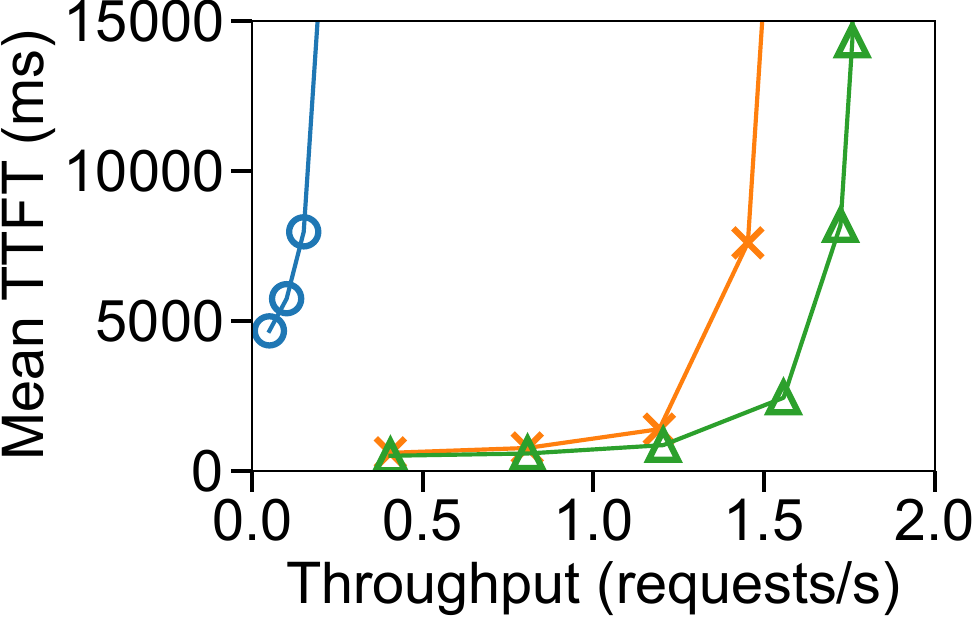}
    \subcaption{Mean TTFT vs. throughput.}
\end{subfigure}
\hfill
\begin{subfigure}[t]{0.213\textwidth}
    \includegraphics[width=\textwidth]{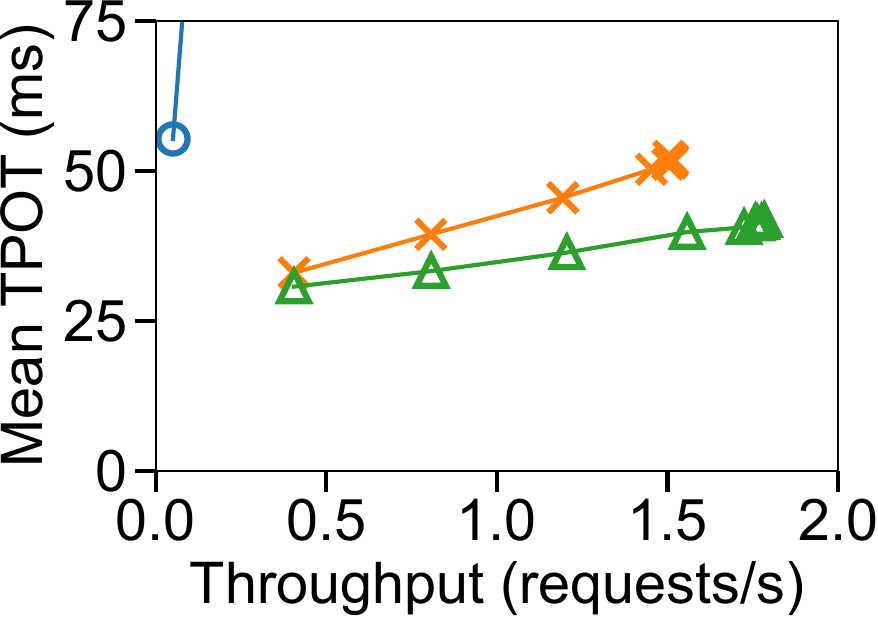}
    \subcaption{Mean TPOT vs. throughput.}
\end{subfigure}
\caption{Load-latency curves for output length = 32 and network bandwidth = 20\,Gbps. The TPOT curves do not show turning points, because TPOT does not include the queueing delay, and therefore does not increase further after the systems become fully loaded.}
\label{fig:eval_curves}
\end{figure}

We first evaluate the end-to-end performance with the Llama-8B model and the NarrativeQA dataset. Figure~\ref{fig:eval_curves} shows how the mean TTFT and TPOT of different systems change with request throughput, when all requests have a fixed output length of 32 and the network bandwidth is limited to 20\,Gbps.

Both \sysname and CacheGen-Async significantly outperform vLLM, which suffers from high latency due to the lack of prefix caching. \sysname achieves 16\% lower unloaded TTFT than CacheGen-Async (502.2ms vs. 600.5ms) for two reasons. First, with its data plane design (\S\ref{sec:chunked_pipeline} and \S\ref{sec:memory_management}), \sysname's fetching and decompression throughput is higher than CacheGen-Async, resulting in lower fetching latency. Second, our use of Deflate as the lossless compression algorithm provides a better compression ratio than CacheGen-Async's arithmetic coding, thus reducing the data transferred over the low-bandwidth network. \sysname also attains 20\% lower loaded TPOT compared to CacheGen-Async (41.8ms vs. 52.0ms). This gap comes again from decompression offloading, as the GPU is now freed to focus solely on model computation (decode), preventing the interference seen in CacheGen-Async. These advantages combined allow \sysname to achieve 18\% higher maximum throughput than CacheGen-Async (1.78\,req/s vs. 1.51\,req/s).

\subsubsection{More Bandwidth and Output Length Settings}
\label{sec:combined_performance}

\begin{figure}[t]
\centering
\begin{subfigure}[t]{0.38\textwidth}
    \includegraphics[width=\textwidth]{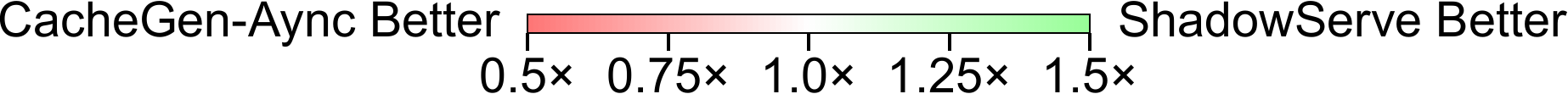}
\end{subfigure}

\vspace{0.4em}

\setcounter{subfigure}{0}
\begin{subfigure}[t]{0.225\textwidth}
    \includegraphics[width=\textwidth]{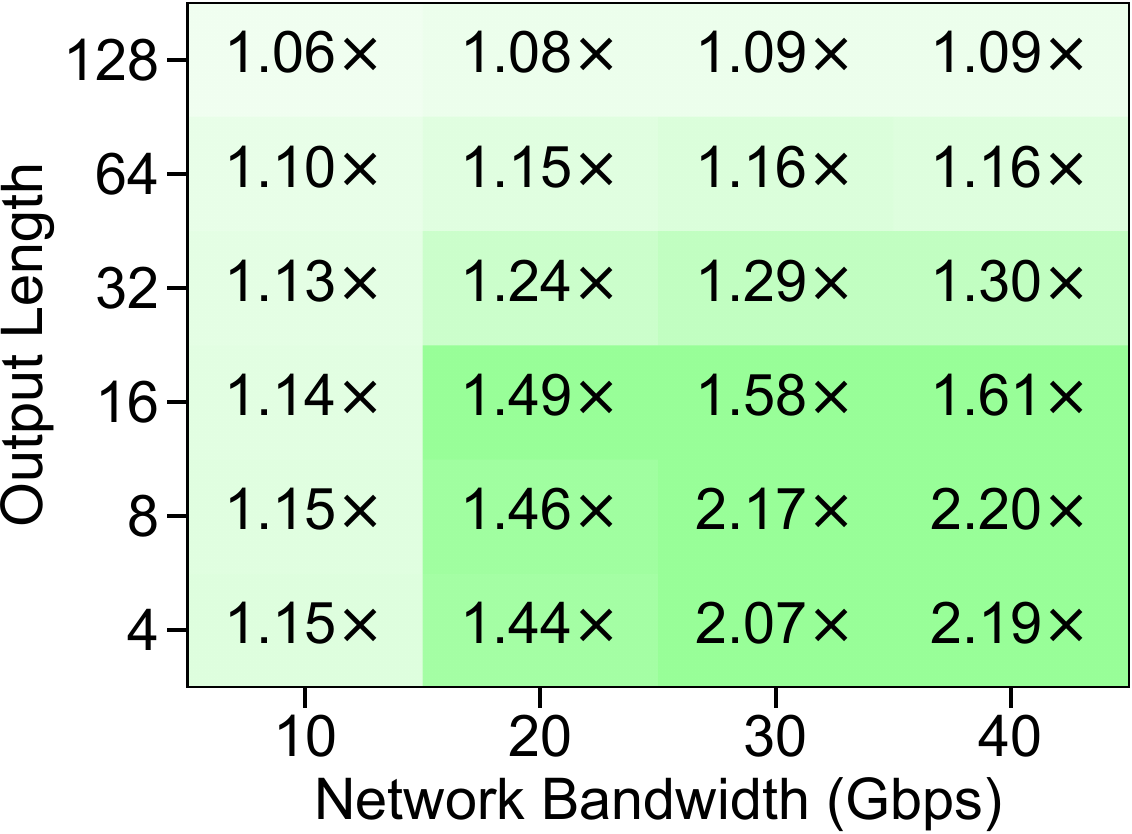}
    \subcaption{Loaded TPOT.}
\end{subfigure}
\hfill
\setcounter{subfigure}{2}
\begin{subfigure}[t]{0.225\textwidth}
    \includegraphics[width=\textwidth]{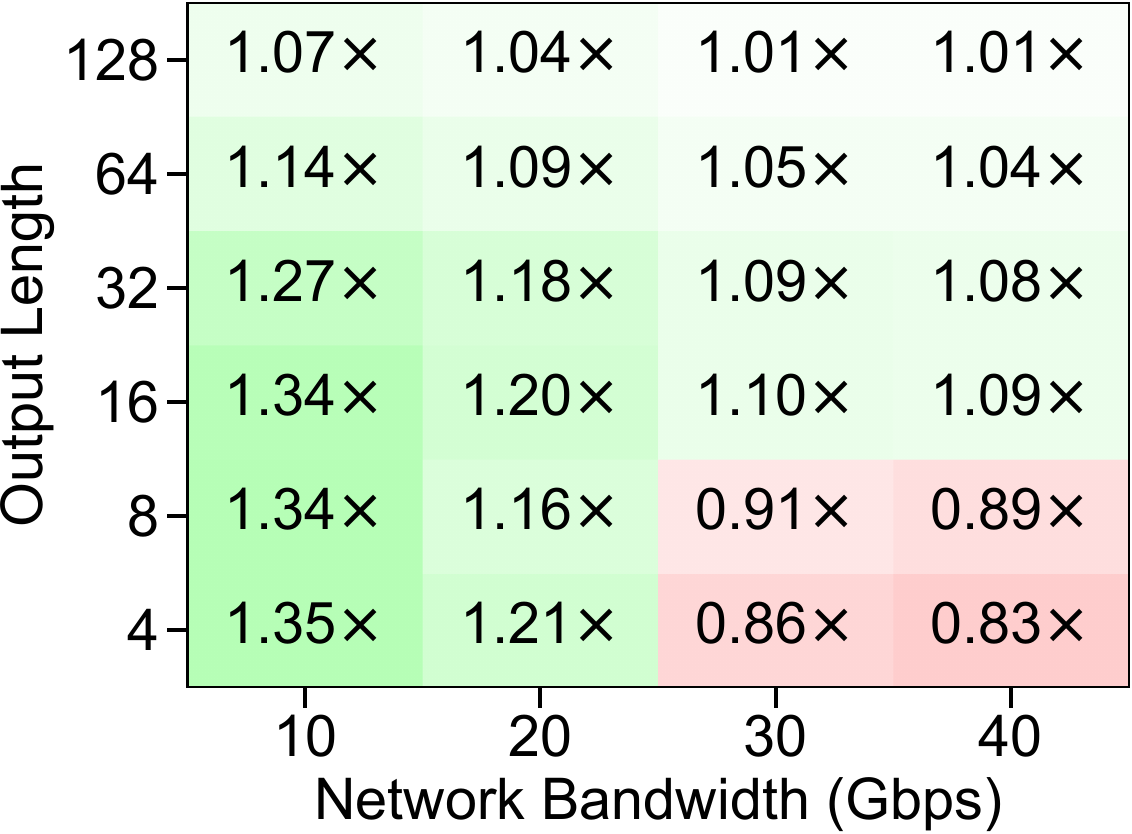}
    \subcaption{Maximum throughput.}
\end{subfigure}

\vspace{0.4em}

\setcounter{subfigure}{1}
\begin{subfigure}[t]{0.25\textwidth}
    \includegraphics[width=\textwidth]{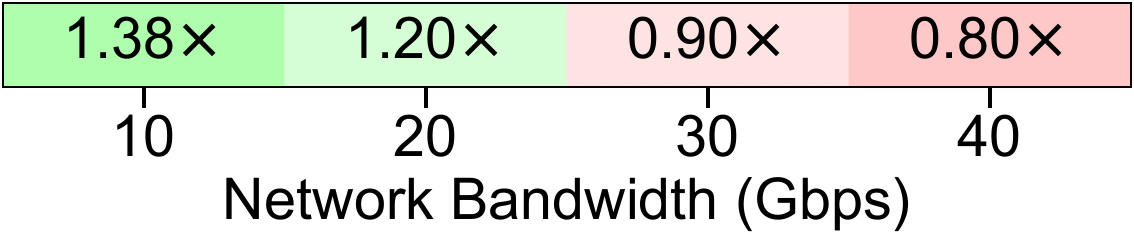}
    \subcaption{Unloaded TTFT.}
\end{subfigure}
\caption{Performance comparison between \sysname and CacheGen-Async across different output lengths and network bandwidths. Unloaded TTFT is not affected by output length as it measures the latency before the first output token.}
\label{fig:combined_performance}
\end{figure}

\begin{figure}[t]
\centering
\includegraphics[width=0.25\textwidth]{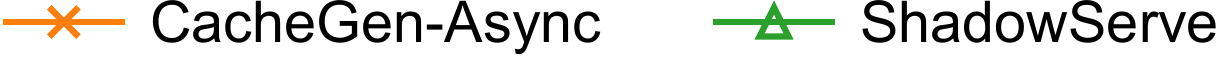}
\vspace{0.2em}

\begin{subfigure}[t]{0.195\textwidth}
    \includegraphics[width=\textwidth]{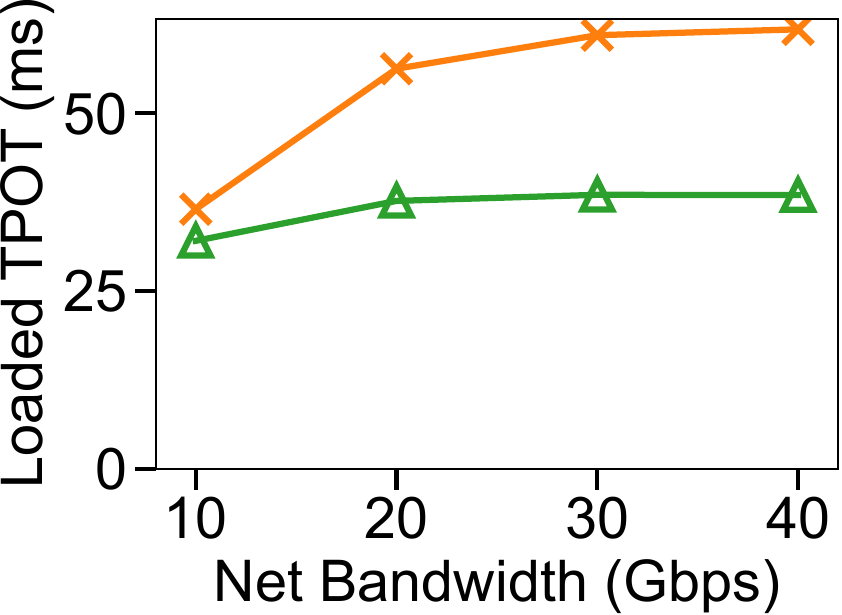}
    \subcaption{Loaded TPOT under output length = 16.}
    \label{fig:bandwidth_vs_tpot_len16}
\end{subfigure}
\hspace{0.5em}
\begin{subfigure}[t]{0.21\textwidth}
    \includegraphics[width=\textwidth]{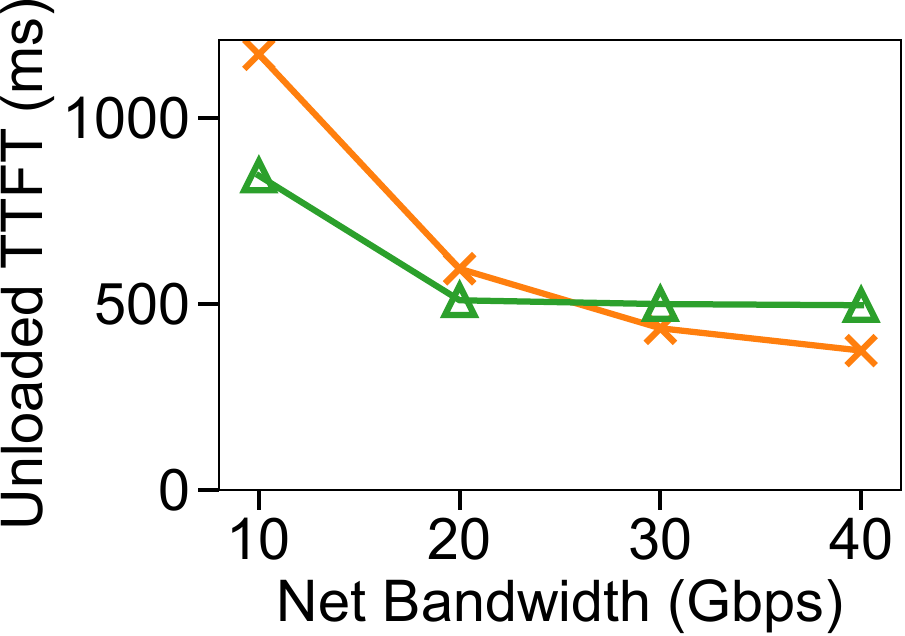}
    \subcaption{Unloaded TTFT.}
    \label{fig:bandwidth_vs_ttft_len16}
\end{subfigure}
\caption{Absolute metrics for \sysname and CacheGen-Async.}
\label{fig:more_settings_curves}
\end{figure}

We extend our evaluation to more network bandwidth and output length configurations to show the entire trade-off space. Figure~\ref{fig:combined_performance} shows the relative improvement of \sysname over CacheGen-Async on maximum throughput, loaded TPOT, and unloaded TTFT, respectively, using Llama-8B and NarrativeQA across a wide range of settings.

\prepara
\para{Loaded TPOT.}
\sysname consistently surpasses CacheGen-Async with 1.06--2.19$\times$ lower loaded TPOT across all output lengths and network bandwidths, demonstrating its ability to eliminate interference between KV cache decompression and model computation. \sysname is able to achieve much less interference because of its much less frequent and less heavyweight kernel launches in the GPU (the per-round scattering kernel discussed in \S\ref{sec:memory_management}) compared with the per-chunk GPU decompression needed in CacheGen-Async.

Zooming in, we show the absolute values for loaded TPOT under 16 output tokens in Figure~\ref{fig:bandwidth_vs_tpot_len16}. CacheGen-Async's loaded TPOT increases dramatically from 36.5ms to 61.8ms (1.7$\times$) as the bandwidth scales up from 10\,Gbps to 40\,Gbps, while \sysname's loaded TPOT only increases slightly from 32.1ms to 38.5ms (1.2$\times$), which leads to \sysname's increasing gain (1.14--1.61$\times$) over CacheGen-Async as network scales. The increase in loaded TPOT in both systems stems from more frequent GPU kernel launches in their pipeline as the network stage becomes shorter for each chunk, leading to more interference with model computation.

\sysname yields less gain in loaded TPOT compared with CacheGen-Async for longer outputs. For example, it only achieves 1.06--1.09$\times$ better for 128 output tokens, compared with up to 2.19$\times$ for 4 output tokens. This occurs because with longer outputs, both systems' performance bottleneck shifts from data plane operations to GPU memory capacity. A request with a long output sequence occupies its KV cache memory region in the GPU for a prolonged decoding phase, stalling new fetches until it completes. During the periods when fetching is stalled, no decompression kernels run on the GPU, which temporarily eliminates the interference that typically penalizes CacheGen-Async. As a result, its TPOT during these compute-only phases becomes nearly optimal, improving its average loaded TPOT and narrowing the performance gap with \sysname.

\prepara
\para{Unloaded TTFT.}
\sysname achieves 1.20--1.38$\times$ lower unloaded TTFT than CacheGen-Async at bandwidths below 20\,Gbps thanks to the better decompression ratio of Deflate, but is 11--24\% higher above 20\,Gbps. Looking more closely at the absolute values shown in Figure~\ref{fig:bandwidth_vs_ttft_len16}, as network bandwidth scales from 10\,Gbps to 40\,Gbps, CacheGen-Async's fetching latency continues to decrease until 40\,Gbps (\eg, an 11\% decrease from 30\,Gbps to 40\,Gbps), while \sysname's fetching latency stops decreasing from 20\,Gbps on, staying at around 500ms. This is caused by the premature network bottleneck in \sysname, as the Bluefield SoC could only achieve 20.6\,Gbps bandwidth because of inter-stage contention in the SmartNIC pipeline, further analyzed in \S\ref{sec:smartnic_perf}.

Nevertheless, these unloaded results assume no interference. Under load, CacheGen-Async will perform worse as its decompression becomes interfered by model computation in the GPU. In addition, based on our measurement, the decompression in CacheGen-Async has a maximum throughput of $\sim$\,32\,Gbps under interference, so CacheGen-Async is already bottlenecked by decompression when the network bandwidth reaches 40\,Gbps. Therefore, above 40\,Gbps, CacheGen-Async's performance stops improving as well.

\prepara
\para{Maximum throughput.}
For shorter output lengths (4 and 8), \sysname achieves 1.16--1.35$\times$ higher throughput in 10 and~20\,Gbps, but becomes 9--17\% worse for 30 and~40\,Gbps, following roughly the same pattern observed in unloaded TTFT. This is because KV cache fetching (TTFT) dominates end-to-end performance for shorter outputs. In addition, shorter outputs result in less model computation (decode), resulting in less GPU interference for CacheGen-Async. Conversely, \sysname beats CacheGen-Async across all bandwidths for longer outputs ($\ge$\,16) due to its lower TPOT, achieving 1.01--1.34$\times$ higher throughput. However, contrary to the pattern observed in loaded TPOT, the advantage of \sysname diminishes as network scales up, due to worse fetching performance (longer TTFT).

\prepara
\para{Summary of trade-off space.}
\sysname consistently excels in TPOT across all configurations due to its interference-free decompression offload. Conversely, CacheGen-Async achieves better TTFT at network bandwidths above 20\,Gbps, as \sysname's fetching speed is bottlenecked by its SmartNIC pipeline. This trade-off dictates the overall throughput: for longer outputs ($\ge$\,16 tokens) or lower bandwidth ($\le$\,20\,Gbps), \sysname delivers higher throughput, while CacheGen-Async pulls ahead when both the output is short and the available bandwidth is high.

Looking beyond our tested settings, we project that for extremely long outputs (>\,128 tokens), the performance of both systems would converge. This is because KV cache fetching will be constantly stalled as the GPU is busy doing many iterations of decode, a setting that neither system optimizes for. Similarly, at bandwidths beyond 40\,Gbps, as both systems are already bottlenecked (CacheGen-Async by GPU decompression, \sysname by the SmartNIC), their relative performance would likely mirror the 40\,Gbps results. At an even higher bandwidth threshold, a baseline without any decompression would likely outperform both, as the latency of decompression itself would outweigh the benefits of reduced data transfer. Ultimately, \sysname is the superior choice for most low-bandwidth prefix caching scenarios especially for TPOT-sensitive workloads, while its main limitation is its raw fetching performance in high-bandwidth networks.

\subsubsection{Different Models and Datasets}

\begin{figure}[t]
\centering
\begin{subfigure}[t]{0.45\textwidth}
    \includegraphics[width=\textwidth]{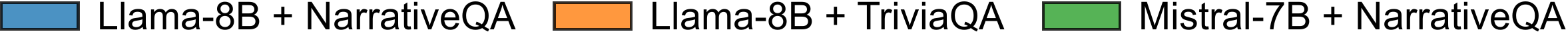}
\end{subfigure}

\vspace{0.4em}

\begin{subfigure}[t]{0.15\textwidth}
    \includegraphics[width=\textwidth]{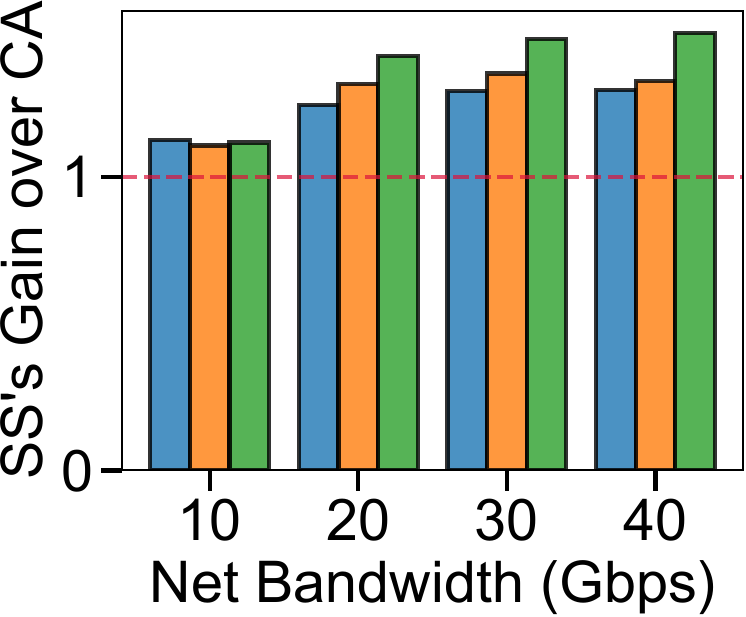}
    \subcaption{Loaded TPOT.}
\end{subfigure}
\hfill
\begin{subfigure}[t]{0.15\textwidth}
    \includegraphics[width=\textwidth]{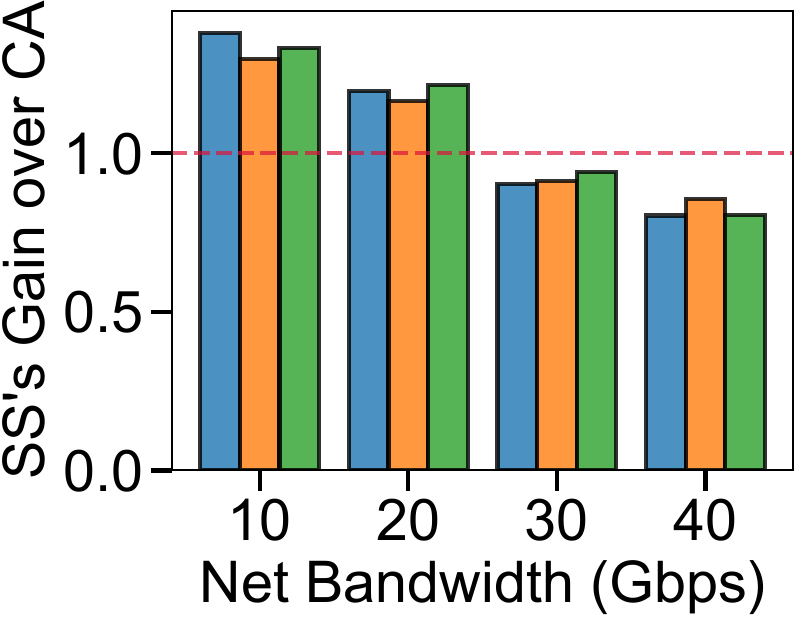}
    \subcaption{Unloaded TTFT.}
\end{subfigure}
\hfill
\begin{subfigure}[t]{0.15\textwidth}
    \includegraphics[width=\textwidth]{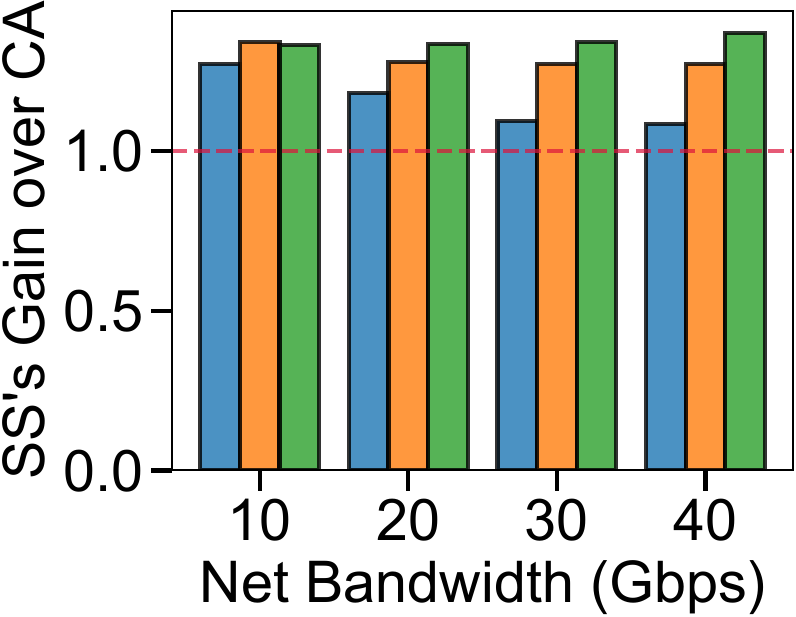}
    \subcaption{Max throughput.}
\end{subfigure}
\caption{\sysname (\shortsysname)'s gain over CacheGen-Async (CA) under more models and datasets, output length = 32.}
\label{fig:more_models_datasets}
\end{figure}

We evaluate \sysname and CacheGen-Async's performance for other models and datasets. Figure~\ref{fig:more_models_datasets} shows \sysname's gain over CacheGen-Async for loaded TPOT, unloaded TTFT, and maximum throughput, with two additional configurations: (Llama-8B + TriviaQA) and (Mistral-7B + NarrativeQA). The output length is fixed at 32 tokens.

\sysname's benefit stays similar across other models and datasets, indicating a very similar trade-off space as discussed in \S\ref{sec:combined_performance}, demonstrating the generality of \sysname's design. For loaded TPOT, \sysname is still better than CacheGen-Async across all settings, with 1.11--1.33$\times$ improvement on (Llama-8B + TriviaQA), and 1.12--1.49$\times$ on (Mistral-7B + NarrativeQA). Like what we observed in \S\ref{sec:combined_performance}, \sysname's gain increases as network bandwidth grows. For unloaded TTFT, \sysname is 1.17--1.38$\times$ better for 10 and~20\,Gbps, and 6--20\% worse for 30 and~40\,Gbps, across all models and datasets. For maximum throughput, \sysname stays ahead by 1.08--1.37$\times$ across all settings.

\subsection{SmartNIC Performance}
\label{sec:smartnic_perf}

\begin{figure}[t]
\centering
\includegraphics[width=0.37\textwidth]{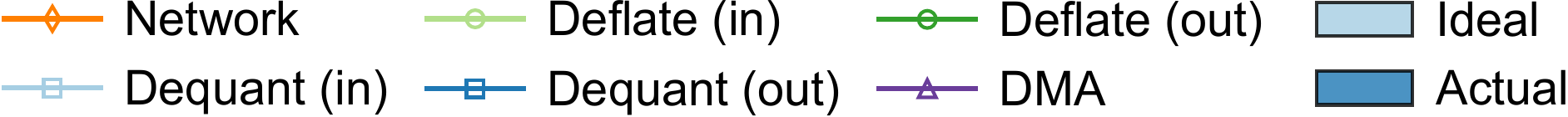}
\vspace{0.4em}

\begin{subfigure}[t]{0.225\textwidth}
    \includegraphics[width=\textwidth]{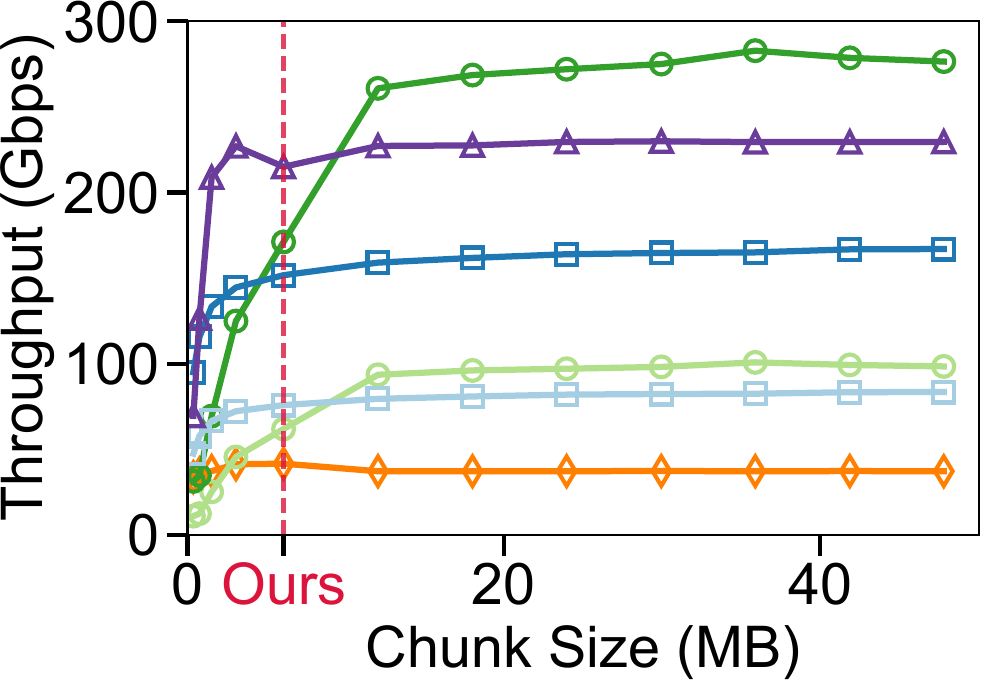}
    \subcaption{Standalone (microbenchmark) performance for each pipeline stage under different chunk sizes.}
    \label{fig:pipeline}
\end{subfigure}
\hfill
\begin{subfigure}[t]{0.225\textwidth}
    \includegraphics[width=\textwidth]{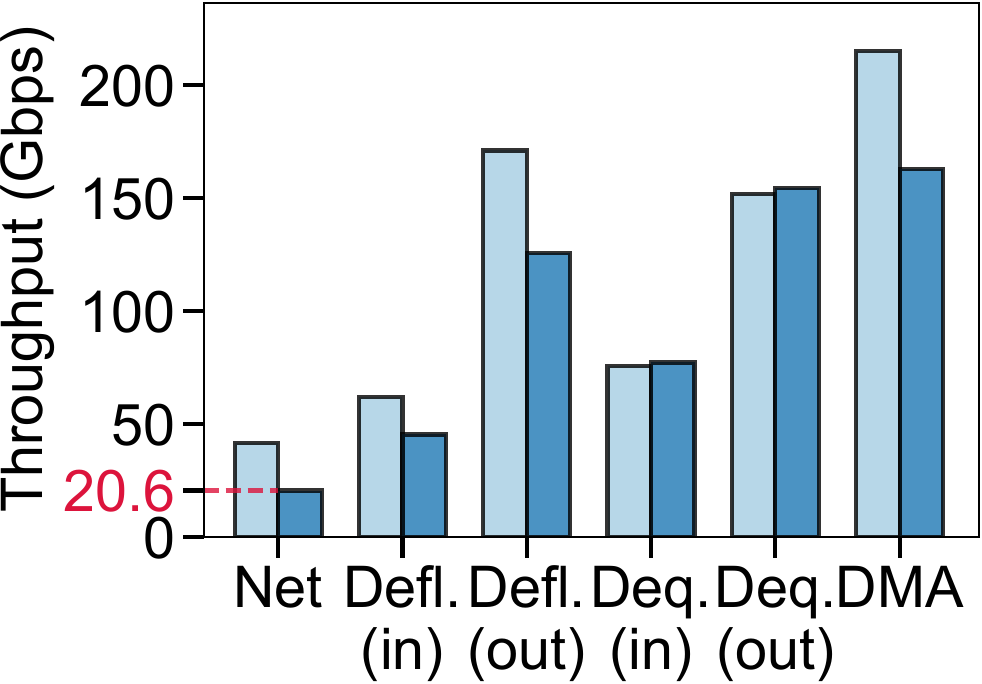}
    \subcaption{Standalone vs. actual performance for each pipeline stage using our chunk size (256 tokens).}
    \label{fig:actual_pipeline}
\end{subfigure}
\caption{SmartNIC microbenchmark and actual performance. There are two throughput numbers for Deflate and dequantization because they have different input and output data sizes. The results show that the pipeline is currently bottlenecked by the network stage.}
\label{fig:micro}
\end{figure}

We characterize the data plane performance on the SmartNIC. Specifically, we examine stages in the chunked pipeline (\S\ref{sec:chunked_pipeline}) and measure their standalone performance with a set of microbenchmarks, as well as their actual performance in the end-to-end system. We then provide an analysis of the performance bottleneck.

\prepara
\para{Standalone throughput of pipeline stages.}
Figure~\ref{fig:pipeline} shows how the throughput of each stage in the data plane's chunked pipeline varies with input chunk size. Standalone means that each stage runs alone on the SmartNIC using its assigned resources in the end-to-end system. For example, we allocate 2 cores to user-space TCP in our implementation (\S\ref{sec:implementation}), so the network curve in Figure~\ref{fig:pipeline} shows the performance of user-space TCP on the two assigned cores without anything else running on the SmartNIC.

As Figure~\ref{fig:pipeline} shows, each operation's throughput increases with chunk size before saturating. The BlueField-3's Deflate accelerator is capable of reaching $\sim$\,280\,Gbps output throughput, and its DMA engine achieves a maximum of 230\,Gbps, which interestingly does not saturate the PCIe 4.0$\times$16 line rate. Dequantization has a lower, yet still substantial, throughput of 83.5/167\,Gbps input/output. In the ideal case, these results show that the pipeline is bottlenecked by the network stage, which saturates at 37.3\,Gbps. This bottleneck dictates our choice of chunk size. The optimal size must be large enough for all other stages to operate above the network's throughput, yet small enough to enable fine-grained pipelining. Therefore, we select a chunk size of 6\,MB (256 tokens).

\prepara
\para{Actual throughput.}
Figure~\ref{fig:actual_pipeline} illustrates the performance degradation when all pipeline stages run together on the SmartNIC. While dequantization maintains its standalone performance, the Deflate and DMA stages experience throughput reductions of 27\% and 24\%, respectively. The most significant impact is on the network stage; already the bottleneck in isolation, its throughput drops by an additional 59\% to 20.6\,Gbps. This degradation is the cause of the suboptimal TTFT observed in \sysname (\S\ref{sec:combined_performance}).

\prepara
\para{SmartNIC bottleneck analysis.}
A common question would be why not increase the number of cores assigned to network to improve its throughput. This is because currently the network operation is not CPU-bound. Indeed, increasing the number of cores assigned to network results in negligible throughput improvement without severely degrading dequantization. We believe BlueField-3's memory subsystem is the cause of the inter-stage contention, because compute resources are already partitioned to each stage (\S\ref{sec:chunked_pipeline}). To validate this, we run the same test as in Figure~\ref{fig:actual_pipeline} but remove all memory accesses in dequantization. In this scenario, per-chunk dequantization latency becomes near-zero, and the network throughput recovers to $\sim$\,30\,Gbps.

We identify two potential sources for this memory contention. The first is the cache hierarchy. The BlueField-3 has 1\,MB, 8\,MB, and 16\,MB L1, L2, L3 caches, respectively~\cite{bf3}, much smaller than the host, and its L3 cache is smaller than the per-chunk working set of the dequantization stage. Our experiments confirm that more than 90\% of L3 cache accesses miss, 40\% of which are caused by dequantization. A smaller chunk size can mitigate this issue, but results in lower operation throughput (Figure~\ref{fig:pipeline}).

However, since the on-chip Deflate and DMA engines bypass the CPU caches, this is unlikely to be the sole factor. Another cause could be memory bandwidth. Although the BlueField-3 is equipped with DDR5 memory offering a high theoretical bandwidth (>\,80\,GB/s), it has only two memory channels~\cite{bf3}, which can lead to head-of-line blocking at the memory controller, leading to suboptimal memory bandwidth. Despite these hardware limitations, our design remains promising. Future SmartNICs can be architected with more memory channels and higher bandwidth. Adding more memory channels is feasible, as Intel Mount Evans IPU includes three channels~\cite{burres2021intel}, and Marvell Octeon DPU supports up to six channels~\cite{marvell-octeon}. Moreover, improving the memory subsystem of the SmartNIC is more cost-effective as Arm SoCs are less expensive than server-class x86 CPUs~\cite{iotcp}.

\subsection{Ablation Studies}
\label{sec:ablation_studies}

\begin{figure}[t]
\centering
\begin{subfigure}[t]{0.31\textwidth}
    \includegraphics[width=\textwidth]{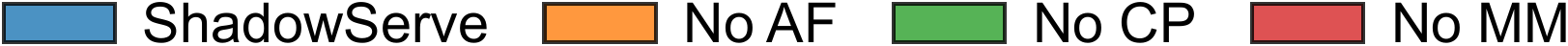}
\end{subfigure}

\vspace{0.3em}

\begin{subfigure}[t]{0.153\textwidth}
    \includegraphics[width=\textwidth]{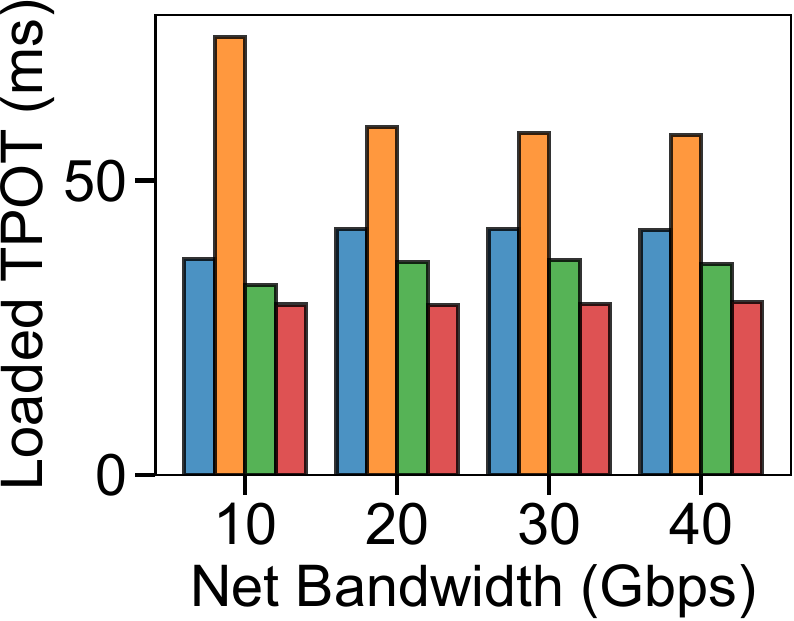}
    \subcaption{Loaded TPOT.}
    \label{fig:ablation_32_tpot}
\end{subfigure}
\hfill
\begin{subfigure}[t]{0.153\textwidth}
    \includegraphics[width=\textwidth]{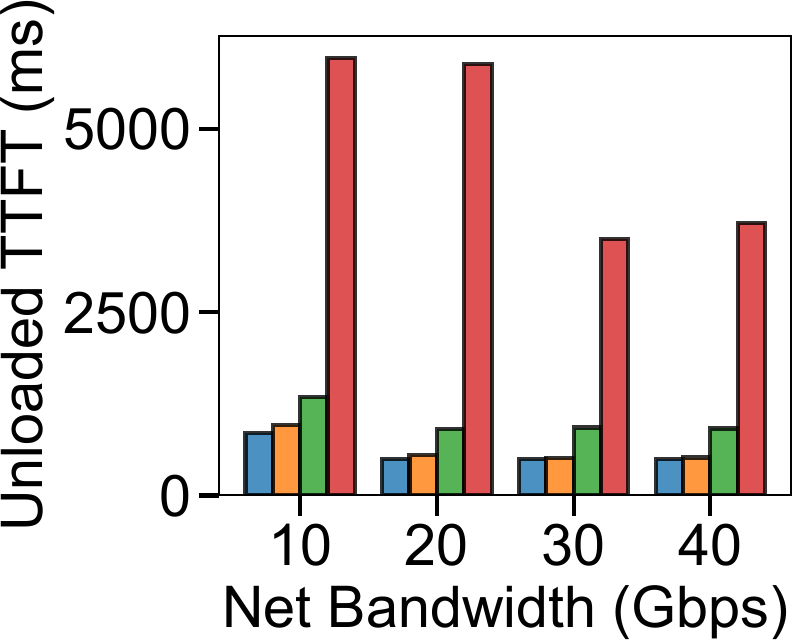}
    \subcaption{Unloaded TTFT.}
\end{subfigure}
\hfill
\begin{subfigure}[t]{0.144\textwidth}
    \includegraphics[width=\textwidth]{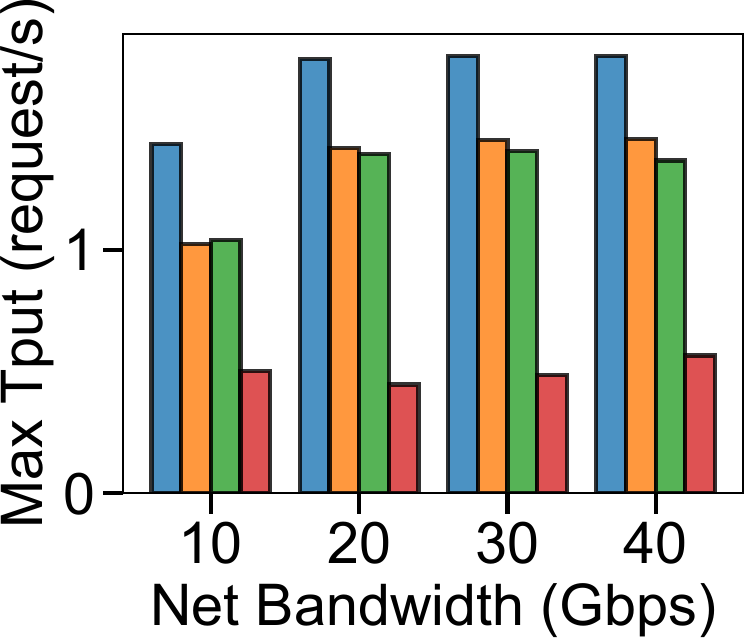}
    \subcaption{Max throughput.}
\end{subfigure}
\caption{Impact of asynchronous fetching (AF), chunked pipeline (CP), and memory management (MM). Output length = 32.}
\label{fig:ablation}
\end{figure}

We evaluate the effectiveness of each component in \sysname (\S\ref{sec:design}) by comparing the performance metrics for the following artifacts, each keeping all other components the same.

\begin{itemize}
    \item \textbf{\sysname.} With all the techniques enabled.
    \item \textbf{No AF.} It turns off asynchronous fetching (\S\ref{sec:asynchronous_fetching}) in \sysname. KV cache fetches block model execution, instead of happening in the background.
    \item \textbf{No CP.} It disables the chunked pipeline (\S\ref{sec:chunked_pipeline}) in the data plane on the SmartNIC. Chunks are processed one by one through the four stages, and each stage uses all resources on the SmartNIC.
    \item \textbf{No MM.} It disables memory management (\S\ref{sec:memory_management}) on the SmartNIC. Instead of pre-allocation and memory pinning, memory buffers are allocated and registered for each chunk on demand during runtime.
\end{itemize}

Figure~\ref{fig:ablation} shows several metrics for the evaluated systems for 32 output tokens. Overall, \sysname achieves 1.2--4.5$\times$ higher maximum throughput, demonstrating the contribution of each technique in \sysname's design. Disabling asynchronous fetching (No AF) does not affect unloaded TTFT, as the fetching itself has roughly the same latency whether it happens in the foreground or background. However, it results in 1.39--2.03$\times$ higher TPOT, as blocking fetches constantly interrupt decodes and cause prolonged stalls between consecutive generated tokens, while \sysname is able to eliminate such stalls by moving KV cache fetching to the background. The gap is larger for lower network bandwidth because fetching takes even longer.

Disabling either the chunked pipeline (No CP) or memory management (No MM) increases unloaded TTFT, as both are critical for data plane performance and low-latency fetching. The impact is particularly severe for No MM, which exhibits a 6.96–11.73$\times$ higher TTFT compared to No CP (1.56–1.86$\times$), due to the high overhead of on-demand memory registration (\S\ref{sec:memory_management}). It might be surprising that disabling the two techniques leads to an improved loaded TPOT (Figure~\ref{fig:ablation_32_tpot}). This is because, as fetching performance deteriorates for No CP and No MM, the scattering kernel in \sysname's data plane (\S\ref{sec:memory_management}) is launched less often in the GPU, leading to slightly less interference and more efficient model computation. As No MM's data plane performance degrades more severely than No CP's, this effect is more pronounced, improving TPOT by 21--31\% for No MM versus 12--14\% for No CP.

\section{Discussion and Future Work}
\label{sec:discussion}

\para{Prefill-decode disaggregation.}
Prefill-decode disaggregation~\cite{splitwise, distserve, kvdirect} assigns prefill and decode to different GPUs, creating significant overhead from transmitting KV caches between them~\cite{distserve, ecoserve, pd-multiplexing, semi-pd}. \sysname can benefit this setting by using its SmartNIC data plane to transparently compress the KV cache on the prefill side and decompress it on the decode side. The network latency can be further hidden with asynchronous sending and receiving (\S\ref{sec:asynchronous_fetching}). Even if only one side is equipped with SmartNICs, single-sided offloading is still beneficial for low-bandwidth settings.

\prepara
\para{Chunked prefill and partial hits.}
Chunked prefill~\cite{sarathi-serve} breaks large prefill jobs into smaller chunks, and puts them in finer-grained mixed batches alongside decodes. \sysname's asynchronous fetching (\S\ref{sec:asynchronous_fetching}) can support this feature. The KV cache manager would modify its batch interception mechanism to inspect the prefill chunks within each mixed batch. When a chunk is eligible for a cache hit, the manager intercepts it and begins the background fetch, using the same process as for full requests. Chunked prefill also enables partial hits. For a partially cached prefill request, the manager fetches the available prefix, and the scheduler computes the remainder as a chunked prefill.

\prepara
\para{Wish list for SmartNICs.}
We highlight the need for more powerful SmartNIC memory subsystems, as \sysname's performance is bottlenecked by the limited cache size and memory bandwidth of current devices (\S\ref{sec:smartnic_perf}). Enhancing these would directly boost the data plane throughput. Additionally, emerging on-path cores (\eg, DPAs on BlueField-3~\cite{benchbf3, scr}) could be a promising alternative to the off-path cores utilized in \sysname, as they can process traffic directly on the data path, bypassing on-chip memory.
%!TEX root = main.tex
\section{Related Work}
\label{sec:relatedwork}

\para{LLM serving.}
Efficient LLM serving requires combining various optimizations. For a single node, fine-grained scheduling~\cite{orca, sarathi-serve}, paged memory management~\cite{vllm}, and operator pipelining~\cite{nanoflow} are proposed. For multiple nodes, PD-disaggregation~\cite{splitwise, distserve} separates prefill and decode phases to different GPUs with elastic parallelism scaling~\cite{loongserve}. Recently, some propose PD-multiplexing on the same device~\cite{pd-multiplexing, semi-pd, bullet} and Attention-FFN disaggregation~\cite{step-3}. All of these systems can leverage prefix caching to avoid recomputation, and thus can benefit from \sysname's design.

\prepara
\para{KV cache compression.}
KV cache compression has been well studied. Lossy algorithms include pruning~\cite{llmlingua, h2o, scissorhands}, low-rank approximation~\cite{deepseek-v2, think}, quantization~\cite{kvquant, gear, kivi}, and many more. Transmission-oriented KV cache compression can be further augmented with lossless algorithms like arithmetic coding~\cite{cachegen}. We emphasize that \sysname does not propose a new KV cache compression algorithm, but offloads existing ones from the host GPU/CPU to the SmartNIC. Therefore, \sysname's design is complementary to new compression algorithms.

\prepara
\para{SmartNIC offloading.}
SmartNICs present a promising approach to reducing host processing overheads~\cite{xenic}. Many applications have been offloaded to SmartNICs, including networking stacks~\cite{zero-nic, flextoe, scr}, file systems and storage~\cite{linefs, leed, gimbal}, distributed transactions~\cite{off-path, xenic}, content delivery~\cite{iotcp}, network functions~\cite{clara, floem, clicknp, azure-smartnic}, and many more~\cite{rpcnic, ipipe, lambda-nic, e3}. \sysname shares the same vision and targets a new realm of applications, \ie, LLM serving, and offloads KV cache compression to the SmartNIC to achieve transparent KV cache fetching.

\prepara
\para{Asynchronous KV cache fetching.}
Recent works also explore asynchronous KV cache fetching. CachedAttention~\cite{cachedattention} overlaps layer-wise pre-loading with computation. However, it needs a dedicated, potentially unbounded pre-loading buffer in the GPU, while \sysname has a bounded GPU memory usage. KVFlow~\cite{kvflow} also employs a proactive prefetching mechanism, but it is designed for agentic workflows with known prefix patterns. In contrast, \sysname supports general serving without this prior knowledge.

\prepara
\para{GPU multitasking.}
GPU multitasking mechanisms like MPS and CUDA Green Context are actively used in LLM serving~\cite{nanoflow, pd-multiplexing, semi-pd, bullet}, but their performance is nondeterministic and workload-dependent as the GPU scheduler is closed-source. \sysname demonstrates their limitations when multiplexing decompression with model computation. Better ways to multiplex tasks on the GPU have been explored, but they either target limited workloads~\cite{antman, zico, pipeswitch, salus, usher} or provide imperfect performance guarantees~\cite{reef, tgs, orion, gpu-multitasking}. In addition, even if GPUs achieve perfect multitasking, offloading is still beneficial as it saves GPU resources for more efficient model serving.

\prepara
\para{Other hardware for tensor compression.}
The latest GPUs are integrating on-chip hardware accelerators for decompression~\cite{blackwell-decompression}, but they are only present in high-end GPUs like NVIDIA Blackwell, and out of scope for low-cost serving. GPU video codecs also yield good compression ratio on KV cache tensors with little loss in accuracy~\cite{vcllm}. However, these codecs have very limited throughput~\cite{vcllm}, and are only present in limited GPU types not suitable for LLM workloads.

%!TEX root = main.tex
\section{Conclusion}
\label{sec:conclusion}

We present \sysname, the first SmartNIC-accelerated prefix caching system that achieves \emph{interference-free} KV cache fetching by offloading the entire data plane to the SmartNIC. \sysname leverages a \emph{chunked pipeline} and a \emph{minimal-copy memory management} scheme to overcome on-device resource limitations and maximize throughput. \sysname achieves up to 2.2$\times$ lower loaded time-per-output-token and up to 1.35$\times$ higher throughput compared to state-of-the-art systems. Our work shows that SmartNICs are a promising and underutilized compute tier for LLM serving infrastructure.

{
\bibliographystyle{plain}
\bibliography{ref}
}

%!TEX root = ./main.tex
\newpage
\appendix

\section{Effect of CUDA Streams}
\label{sec:more_baselines}

\begin{figure}[t]
\centering
\includegraphics[width=0.45\textwidth]{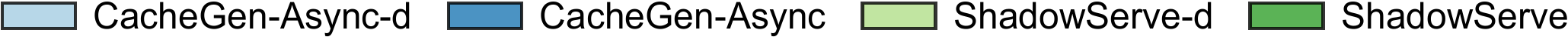}

\vspace{0.5em}

\begin{subfigure}[t]{0.33\textwidth}
    \includegraphics[width=\textwidth]{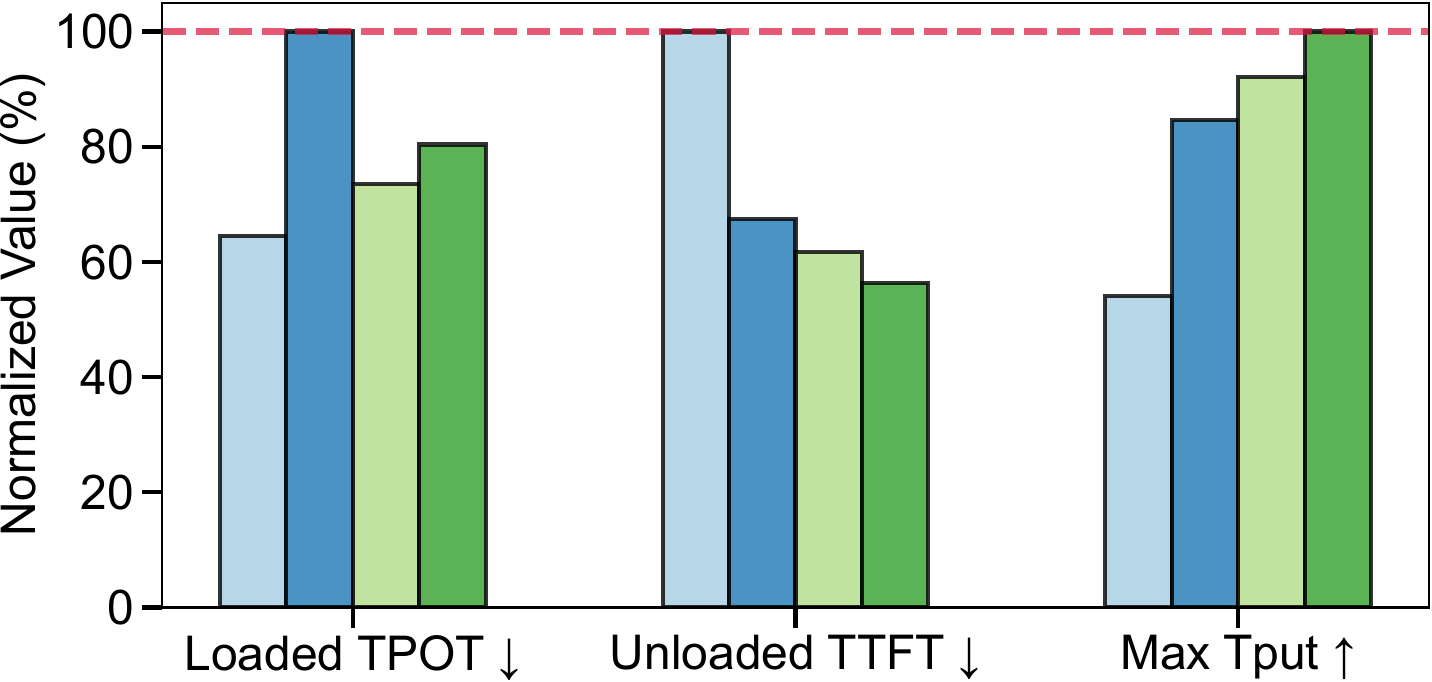}
\end{subfigure}
\caption{Performance of more baselines for output length = 32 and network bandwidth = 20\,Gbps.}
\label{fig:more_baselines}
\end{figure}

Both CacheGen-Async and \sysname use two custom CUDA streams in the GPU for multitasking. We evaluate two additional baselines, \sysname-d and CacheGen-Async-d, that use the default stream for model computation. Figure~\ref{fig:more_baselines} shows normalized metrics for the systems under 32 output tokens and 20\,Gbps bandwidth. For both systems, moving model computation to the default stream leads to lower loaded TPOT (35\% lower for CacheGen-Async-d and 8\% for \sysname-d) and higher unloaded TTFT (48\% and 9\%, respectively), creating interesting new points in the trade-off space. This is because kernels in the default stream (model computation) are prioritized over those in the custom stream (decompression for CacheGen-Async, and scattering for \sysname) in this setting. The effect is much more pronounced for CacheGen-Async, as it launches much more kernels in the custom stream. CacheGen-Async-d even achieves 20\% lower loaded TPOT than \sysname. However, its prolonged TTFT leads to 46\% lower throughput.

We note that this effect caused by the default stream is non-deterministic, as the GPU scheduler is closed-source. For example, we observe the opposite behavior in Figure~\ref{fig:interference_dequant}, where using a custom stream for model computation yields better decode performance than using the default stream.

\end{document}